\documentclass[12pt]{elsarticle}
\usepackage{amssymb}
\usepackage{amsmath}
\usepackage{multirow}

\newtheorem{theorem}{Theorem}[section]
\newtheorem{lemma}[theorem]{Lemma}
\newtheorem{proposition}[theorem]{Proposition}
\newtheorem{corollary}[theorem]{Corollary}
\newtheorem{rem}[theorem]{Remark}

\journal{Computational Statistics \& Data Analysis}

\newcommand{\tr}{\operatorname{tr}}
\def\e{\mathrm e}
\def\E{\mathbb E}
\graphicspath{{.}{Figures/}}

\begin{document}

\begin{frontmatter}

\title{A general procedure to combine estimators}

\author[nantes,inria]{F. Lavancier\corref{cor1}}
\ead{lavancier@univ-nantes.fr}
\cortext[cor1]{Corresponding author}

\author[nantes]{P. Rochet}
\ead{rochet@univ-nantes.fr}

\address[nantes]{University of Nantes, Laboratoire de Math\'ematiques Jean Leray, 2 rue de la Houssini\`{e}re, 44322 Nantes, France}
\address[inria]{Inria, Centre Rennes  Bretagne Atlantique, Campus universitaire de Beaulieu
35042 Rennes, France}

\begin{abstract}
A general method to combine several estimators of the same quantity is investigated. In the spirit of model and forecast averaging, the final estimator is computed as a weighted average of the initial ones, where the weights are constrained to sum to one. In this framework, the optimal weights, minimizing the quadratic loss, are entirely determined by the mean squared error matrix of the vector of initial estimators. The averaging estimator is built using an estimation of this matrix, which can be computed from the same dataset. A non-asymptotic error bound on the averaging estimator is derived, leading to asymptotic optimality under mild conditions on the estimated mean squared error matrix. This method is illustrated on standard statistical problems in parametric and semi-parametric models where the averaging estimator outperforms the initial estimators in most cases.
\end{abstract}

\end{frontmatter}


\section{Introduction}\label{intro}
We are interested in estimating a parameter $\theta$ in a statistical model, based on a collection of preliminary estimators $T_{1},...,T_{k}$. In general, the relative performance of each estimator depends on the true value of the parameter, the sample size, or other unknown factors, in which case deciding in advance what method to favor can be difficult. This situation occurs in numerous problems of modern statistics like forecasting or non-parametric regression, but it remains a major concern even in simple parametric problems. In this paper, we study a general methodology to combine linearly several estimators in order to produce a final single better estimator. \\

The issue of dealing with several possibly competing estimators of the same quantity has been extensively studied in the literature these past decades. One of the main solution retained is to consider a weighted average of the $T_i$'s. The idea of estimator averaging actually goes back to the early 19th century with Pierre Simon de Laplace \cite{laplace},  who was interested in finding the best combination between the mean and the median to estimate the location parameter of a symmetric distribution, see the discussion in \cite{stiegler}. More generally, the solution can be expressed as a linear combination of the initial estimators
\begin{equation}\label{ag0}\hat \theta_\lambda = \lambda^\top  \mathbf T = \sum_{i=1}^k \lambda_i T_i,\end{equation}
for $\lambda$ a vector of weights lying in a subset $\Lambda$ of $\mathbb R^k$ and $\mathbf T=(T_1,\dots,T_k)^\top$. A large number of statistical frameworks fit with this description. For example, model selection can be viewed as a particular case for $\Lambda$ the set of vertices. Similarly, convex combinations corresponds to the simplex $\Lambda = \{ \lambda: \sum_{i=1}^k \lambda_i = 1, \lambda_i \geq 0 \}$ while linear combinations to $\Lambda = \mathbb R^k$. Another well-used framework consists in relaxing the positivity condition of convex combination, corresponding to the set $\Lambda = \{ \lambda: \sum_{i=1}^k \lambda_i = 1\}$.\\

Estimator averaging has received a particular attention for prediction purposes. Ever since the paper of Bates and Granger \cite{bates1969combination}, dealing with forecast averaging for time series, the literature on this subject has greatly developed, see for instance \cite{elliott2011averaging,timmermann2006forecast} in econometrics and \cite{Cesa-Lugosi} in machine learning. In this framework, the parameter $\theta$ represents the future observation of a series to be predicted and $\mathbf T=(T_1,...,T_k)$ a collection of predictors. Averaging methods have also been widely used for prediction in a regression framework. In this case, $\theta$ is the response variable to predict given some regressors, and $\mathbf T$ is a collection of models output. These so-called model averaging procedures are shown to provide good alternatives to model selection for parametric regression, see \cite{moral2010model} for a survey. Model averaging has been studied in both Bayesian \cite{raftery1997bayesian,wasserman2000bayesian} and frequentist  contexts \cite{buckland1997, hansen2007least, hjort2003frequentist}. In closed relation, functional aggregation deals with the same problem in  non-parametric regression  \cite{MR2280619,MR3059085,nemirovski2000,MR3225246,yang2004aggregating}. Aggregation methods have also been extensively studied for density estimation, as an alternative to classical bandwidth selection methods \cite{MR2397610,Catoni2004,rigollettsybakov,MR1762904}. \\

In \cite{hansen2007least}, Hansen introduced a least squares model average estimator, in the same spirit as the forecast average estimator proposed in \cite{bates1969combination}. Loosely speaking, this estimator  aims to mimic the oracle, defined as the linear combination $\hat \theta_\lambda$ that minimizes the quadratic loss $\mathbb E (\hat \theta_\lambda - \theta)^2$, under the constraint on the weights $\sum_{i=1}^k \lambda_i = 1$. Under this constraint, the oracle expresses in terms of the mean squared error matrix $\Sigma$ of $\mathbf T$. The averaging estimator is then defined by replacing $\Sigma$ by an estimator $\hat \Sigma$. 

The main objective of this paper is to apply the latter idea to classical estimation problems, not restricted to prediction. Although it can be applied to non-parametric models,  our procedure is essentially designed for parametric or semi-parametric models, where  the number $k$ of available estimators is small compared to the sample size $n$ and does not vary with $n$. The procedure works well in these situations because the estimation of $\Sigma$ can be carried out efficiently by standard methods (e.g. plug-in or Monte-Carlo), and does not require the tuning of extra parameters. While it recovers some results of \cite{MR0107925, halperin1961almost, MR0264806} and more recently \cite{MR2126899} on estimator averaging for the mean in a Gaussian model, the method applies to a wide range of statistical models. It is implemented in Section~\ref{simus} on four other examples. In the first one, $\theta$ represents the position of an unknown distribution, which can be estimated by both the empirical mean and median, as initially addressed by P. S. de Laplace in \cite{laplace}. In the second example, $\theta$  is  the two-dimensional parameter of a Weibull distribution,  for which several competing estimators exist. In the third one, we consider a stochastic process, namely the Boolean model, that also depends on a two-dimensional parameter and we apply averaging to get a better estimate.  The fourth example deals with estimation of a quantile from the combination of a non-parametric estimator and  possibly misspecified parametric estimators. \bigskip

An important contribution  of our approach is to include the case where several parameters $\theta_1,\dots,\theta_d$ have to be estimated, and a collection of estimators is available for each of them. In order to fully exploit the available information to estimate say $\theta_1$, it may be profitable to average all estimators, including those designed for $\theta_j$, $j\neq 1$. We show that a minimal requirement is that the weights associated to the latter estimators sum to 0, while the weights associated to the estimators of $\theta_1$ sum to one (additional  constraints on the weights can also be added as discussed in Section~\ref{sec:set}). To our knowledge, estimator averaging including estimators of other parameters is a new idea. Our simulation study shows that it can produce conclusive results in some specific situations such as the Boolean model treated in the third example of Section \ref{simus}. \\

From a theoretical point of view, we provide an upper bound on the deviation of the averaging estimator to the oracle. Our result is non-asymptotic and involves the error to the oracle for the actual event, in contrast with usual criteria based on expected loss functions.  In particular, our result strongly differs from classical oracle inequalities derived in the literature on  aggregation for non-parametric regression  \cite{MR2280619,MR3059085,MR1792783,yang2004aggregating} or density estimation \cite{MR2397610,Catoni2004,rigollettsybakov,MR1762904}.  
Moreover, we deduce that under mild assumptions, our averaging estimator behaves asymptotically as the oracle, generalizing the  asymptotic optimality result proved by Hansen and Racine  \cite{hansen2012jackknife} in the frame of model averaging where $\Sigma$ is estimated by jackknife. Our result applies in particular if $\sqrt n (\mathbf T - \theta)$ converges in quadratic mean to a Gaussian law and a consistent estimator of the asymptotic covariance matrix is available, though these conditions are far from being necessary. This situation makes it possible to construct an asymptotic confidence interval based on the averaging estimator, the length of which  is necessarily smaller than all confidence intervals based on the initial estimators.

\bigskip

The remainder of the paper is organized as follows. The averaging procedure is detailed in Section~\ref{construction}, where we give some examples for the choice of the set of weights $\Lambda$, or equivalently of the constraints followed by the weights, and we detail some methods for the estimation of $\Sigma$. In Section~\ref{theory} we prove a non-asymptotic bound on the error to the oracle and discuss the asymptotic optimality of the averaging estimator. 
Section~\ref{simus} is devoted to some numerical applications where we show that the method performs almost always better than the best estimator in the initial collection $\mathbf T$ when the model is well-specified, and is quite robust to misspecification problems. Proofs of our results are postponed to the Appendix.

\section{The averaging procedure}\label{construction} 

The method is different whether it is applied to one parameter or several. For ease of comprehension, we first present the averaging procedure for one parameter, which follows the idea introduced in \cite{bates1969combination} for forecast averaging, though our choice of the set of weights $\Lambda$ may be different. We then introduce a generalization of the procedure for averaging several parameters simultaneously.  Finally, we discuss in Sections~\ref{sec:set} and \ref{sec:mse}  the choice of $\Lambda$ and the construction of $\hat\Sigma$.

\subsection{Averaging for one parameter}\label{ag_oneparameter}
Let $\mathbf T = (T_1,...,T_k)^\top$ be a collection of estimators of a real parameter $\theta$. We search for a decision rule that combines suitably the $T_i$'s to provide a unique estimate of $\theta$. A widely spread idea is to consider linear transformations
$$ \hat \theta_\lambda = \lambda^\top \mathbf T, \ \lambda \in \Lambda, $$
where $\lambda^\top $ denotes the transpose of $\lambda$ and $\Lambda$ is a given subset of $\mathbb R^k$. In this linear setting, a convenient way to measure the performance of $\hat \theta_\lambda$ is to compare it to the \textit{oracle} $\hat \theta^*$, defined as the best linear combination $\hat \theta_\lambda$ obtained for a non-random vector $\lambda \in \Lambda$. Specifically, the oracle is the linear combination $\hat \theta^* = {\lambda^* }^\top  \mathbf T$ minimizing the mean squared error (MSE), i.e. 
$$ \lambda^* = \arg \min_{\lambda \in \Lambda} \ \mathbb E (\lambda^\top  \mathbf T - \theta)^2.$$
Of course, $\lambda^*$ is unknown in practice and needs to be approximated by an estimator, say $\hat\lambda$. \\

The performance of the averaging procedure highly relies on the choice of the set $\Lambda$. Indeed, choosing a too large set $\Lambda$ might increase the accuracy of the oracle but make it difficult to estimate $\lambda^*$. On the contrary, a too small set $\Lambda$ might lead to a poorly efficient oracle but easy to approximate. Therefore, a good balance must be found for the oracle to be both accurate and reachable. In this purpose, a choice proposed in \cite{bates1969combination} and widely used in the averaging literature is to impose the condition $\lambda^\top \mathbf 1 =1$ on the weights, where $\mathbf 1$ denotes the unit vector $\mathbf 1 = (1,...,1 )^\top $. We explain in Section~\ref{sec:set} why this condition  is minimal for the efficiency of the averaging procedure when $\theta\in\mathbb R$.
Moreover, if one wants to impose additional constraints on the weights, such as positivity for instance, the method proposed in this paper allows one to consider as the constraint set $\Lambda$, any non-empty closed subset of
$$\Lambda_{\max} := \{ \lambda \in \mathbb R^k: \lambda^\top  \mathbf 1 = 1 \}.$$
Some examples of constraint sets $\Lambda$ are discussed in Section~\ref{sec:set}.

We assume that the initial estimators have finite order-two moments and $1,T_1,...,T_k$ are linearly independent so that the Gram matrix
$$ \Sigma  = \mathbb E \big[ (\mathbf T - \theta \mathbf 1) (\mathbf T - \theta \mathbf 1 )^\top \big] $$
is well defined and non-singular. From  the identity $\lambda^\top   \mathbf 1 = 1$, we see that the optimal weight $\lambda^*$ defining the oracle $\hat \theta^* = {\lambda^*}^\top  \mathbf T$ writes 
$$ \lambda^* = \arg \min_{\lambda \in \Lambda} \ \mathbb E (\lambda^\top  \mathbf T - \theta)^2 = \arg \min_{\lambda \in \Lambda} \ \lambda^\top  \Sigma \lambda.  $$
Remark that the assumptions made on $\Lambda$ ensure the existence of a minimizer. If $\Lambda$ is convex, the solution is unique, otherwise we agree that $\lambda^*$ refers to one of the  minimizers. In the particular important example where $\Lambda = \Lambda_{\max}$, we get the explicit solution
$$ \lambda^*_{\max} = \frac{\Sigma^{-1} \mathbf 1}{\mathbf 1^\top  \Sigma^{-1} \mathbf 1}, $$
considered for instance in \cite{bates1969combination}, \cite{elliott2011averaging} or \cite{hansen2012jackknife}. In practice, the MSE matrix $\Sigma$ is unknown and has to be approximated by some estimator $\hat\Sigma$ to yield the averaging estimator $\hat \theta=\hat\lambda^\top \mathbf T$, where 
$$\hat\lambda =  \arg \min_{\lambda \in \Lambda} \ \lambda^\top  \hat\Sigma \lambda.$$
Possible methods to  construct $\hat \Sigma$ are discussed in Section~\ref{sec:mse}. While it may seem paradoxical  to shift our attention from $\theta$ to the less accessible $\Sigma$, the effectiveness of the averaging process can be explained by a lesser sensibility to the errors on $\hat \Sigma$. As a result, the averaging estimator improves on the original collection as soon as we are able to build $\hat \Sigma$ sufficiently close from the true value, without stronger requirement such as consistency. On the contrary, the chances of considerably deteriorating the estimation of $\theta$ are expected to be small due to the smoothing effect of averaging.

\subsection{Averaging for several parameters}\label{sec:agseveral}
We now discuss a generalization of the method that deals with several parameters simultaneously. Let $\theta=(\theta_{1}, \dots , \theta_{d})^\top  \in  \mathbb R^d$ and assume we have access to a collection of estimators $\mathbf T_j$ for each component $\theta_j$. For sake of generality we allow the collections $\mathbf T_1, \dots, \mathbf T_d$ to have different sizes $k_1,\dots, k_d$ with $k_j \geq 1$. So, let $\mathbf T_1 \in \mathbb R^{k_1}, \dots , \mathbf T_d \in \mathbb R^{k_d}$ and set $\mathbf T = (\mathbf T_1 ^\top , \dots , \mathbf T_d^\top )^\top  \in \mathbb R^k$, with $k = \sum_{j=1}^d k_j \geq d$. We consider averaging estimators of $\theta$ of the form
$$ \hat \theta_\lambda = \lambda^\top  \mathbf T \in \mathbb R^d,   $$
where here, $\lambda$ is a $k \times d$ matrix. In order to make the oracle more accessible, we impose some restrictions on the set of authorized values for $\lambda$. In this purpose, define the matrix
$$\operatorname{J} = \left( \begin{array}{cccc} \!\! \mathbf 1_{k_1} \!\! & 0 & \dots & 0 \\ 0 &  \!\! \mathbf 1_{k_2} \!\!  & \ddots & \vdots  \\ \vdots & \ddots & \ddots & 0 \\ 0 & \dots & 0 &  \!\! \mathbf 1_{k_d} \!\! \end{array} \right) \in \mathbb R^{k \times d}, $$
where $\mathbf 1_{k_j}$ is the vector composed of $k_j$ ones (we simply denote it $\mathbf 1$ in the sequel to ease notation). We consider the maximal constraint set 
\begin{equation}\label{lambdamax} \Lambda_{\max} = \{ \lambda \in \mathbb R^{k \times d}: \lambda ^\top  \operatorname{J}= \text I \}, \end{equation}
with $\text I$ the identity matrix. Let $\underline{\lambda}_j \in \mathbb R^k$ denote the $j$-th column of $\lambda \in \mathbb R^{k \times d}$. For each component $\theta_{j}$, the average is given by
$$ \hat \theta_{\lambda,j} = \underline{\lambda}_j^\top  \mathbf T = \underline{\lambda}_{j,1}^\top  \mathbf T_1 + \dots + \underline{\lambda}_{j,d}^\top  \mathbf T_d,   $$
where $\underline{\lambda}_j = (\underline{\lambda}_{j,1}^\top, \dots , \underline{\lambda}_{j,d}^\top)^\top$ with $\underline{\lambda}_{j,\ell} \in \mathbb R^{k_\ell}$, $\ell=1,\dots,d$.
 Imposing that $\lambda \in \Lambda_{\max}$ means that for any $ j=1,\dots,d$,
\begin{equation}\label{cond}  \underline{\lambda}_{j,\ell}^\top  \mathbf 1 = \left\{ \begin{array}{ll} 0 & \text{if} \ \  \ell \neq j \\ 1 & \text{if} \ \ \ell =j. \end{array} \right. \end{equation}
This condition does not rule out using the entire collection $\mathbf T$ to estimate each component $\theta_{j}$, although the weights $ \underline{\lambda}_{j,\ell}$ do not satisfy the same constraints depending on the relevance of $\mathbf T_\ell$. While it may seem more natural to impose that only $\mathbf T_j$ is involved in the estimation of $\theta_{j}$ (and this can be made easily through an appropriate choice of $\Lambda\subset\Lambda_{\max}$, letting $\underline{\lambda}_{j,\ell} = 0$ for $\ell \neq j$), allowing one to use the whole set $\mathbf T$ to estimate each component enables to take into account possible dependencies, which may improve the results. Finally, remark that if the collections $\mathbf T_j$ are uncorrelated, the two frameworks are identical.\\

From a technical point of view, the condition $\lambda^\top \operatorname{J} = \text I$ is imposed to have the equality $ \lambda^\top  \mathbf T - \theta = \lambda^\top  (\mathbf T - \operatorname{J} \theta)$, which is used to derive the optimality result of Theorem \ref{slutsky2multi} in Section \ref{errorbound}. Letting $\Vert . \Vert$ denote the usual Euclidean norm on $\mathbb R^d$, the mean squared error then becomes, using the classical trick of switching trace and expectation,
\begin{equation}\label{traceexp} \mathbb E \Vert \lambda^\top  \mathbf T - \theta \Vert^2 = \mathbb E \big[ \tr \big[ (\mathbf T - \operatorname{J} \theta)^\top  \lambda \lambda^\top (\mathbf T - \operatorname{J} \theta) \big] \big]  = \tr(\lambda^\top  \Sigma \lambda), \end{equation}
where $\Sigma = \mathbb E \big[ (\mathbf T - \operatorname{J} \theta )(\mathbf T - \operatorname{J} \theta)^\top \big] \in \mathbb R^{k \times k}$. Here again, we assume that $\Sigma$ exists and is non-singular. \\

Ideally, one would want to minimize the matrix mean squared error $ \mathbb E \big[ (\lambda^\top  \mathbf T - \theta) (\lambda^\top  \mathbf T - \theta)^\top \big] = \lambda^\top \Sigma \lambda$, and not only its trace, for $\lambda^*$ to satisfy the stronger property 
\begin{equation}\label{matrixoracle} \forall  \lambda \in \Lambda, \ \ \ \lambda^\top \Sigma \lambda - \lambda^{*\top} \Sigma \lambda^* \text{ is non-negative definite}.
\end{equation}
Notice however that comparing $\lambda$ and $\lambda^*$ according to this criterion is not always possible since it involves a partial order relation over the matrices and a solution to \eqref{matrixoracle} might not exist. By considering its trace, we are guaranteed to reach an admissible solution, that is, a solution for which no other value is objectively better in this sense. In particular, minimizing $\lambda \mapsto \tr(\lambda^\top \Sigma \lambda)$ reaches the unique solution $\lambda^*$ of \eqref{matrixoracle} whenever one exists. This occurs for instance for $\Lambda = \Lambda_{\max}$, as we point out in Section \ref{sec:set}. \\

The simultaneous averaging process for several parameters generalizes the procedure presented in Section~\ref{ag_oneparameter}. In fact, averaging for one parameter just becomes the particular case with $d=1$. Given a subset $\Lambda \subseteq\Lambda_{\max}$, we define the oracle as the linear transformation $\hat \theta^* = {\lambda^*}^\top  \mathbf T$ with
\begin{equation}\label{oracle} \lambda^* = \arg \min_{\lambda \in \Lambda} \ \mathbb E \Vert \lambda^\top  \mathbf T - \theta \Vert^2 = \arg \min_{\lambda \in \Lambda} \ \tr(\lambda^\top  \Sigma \lambda) . \end{equation}

\noindent Finally, assuming we have access to an estimator $\hat \Sigma$ of $\Sigma$, see Section~\ref{sec:mse}, we define the averaging estimator as $\hat \theta = \hat \lambda^\top  \mathbf T$ where
\begin{equation}\label{hattheta}  \hat \lambda = \arg \min_{\lambda \in \Lambda} \ \operatorname{tr}( \lambda^\top  \hat \Sigma \lambda) . \end{equation}
If $\lambda^\top  \Sigma \lambda$ is well approximated by $\lambda^\top  \hat \Sigma \lambda$ for $\lambda \in \Lambda$, we can reasonably expect the average $\hat \theta$ to be close to the oracle $\hat \theta^*$, regardless of the possible dependency between $\hat \Sigma$ and $\mathbf T$.

\subsection{Choice of the constraint set}\label{sec:set}
The constraint set plays a crucial part in the averaging procedure. As discussed in the previous sections, an appropriate choice of $\Lambda$ must take into account both the accuracy of the oracle and the ability to estimate $\lambda^*$. Writing the estimation error as
\begin{equation}\label{eq1}  \hat \theta - \theta =  \hat \theta^* - \theta + (\hat \lambda - \lambda^*)^\top  \mathbf T, \end{equation}
a good rule of thumb is to choose a set $\Lambda$ as large as possible, but for which the residual term $(\hat \lambda - \lambda^*)^\top  \mathbf T$ can be made negligible compared to the error of the oracle $\hat \theta^* - \theta$. Without restrictions on $\lambda$, Equation \eqref{eq1} suggests that we would have to estimate the optimal combination $\lambda^*$ more efficiently than we can estimate $\theta$. This condition can be reasonably expected for high-dimensional parameters $\theta$, e.g. for non-parametric regression or density estimation, and linear aggregation has indeed been shown to be particularly well adapted to these frameworks, see for instance 
\cite{MR2280619,MR2397610,MR3059085,nemirovski2000,rigollettsybakov,MR3225246,MR1762904}. Nevertheless, hoping for $\hat \lambda - \lambda^*$ to be negligible compared to $\hat \theta^* - \theta$ seems rather unrealistic when $\theta$ is vector-valued, especially given that the optimal combination needs to be estimated for every component $\theta_j$. Even in the simple case $d=1$, the oracle obtained over $\Lambda=\mathbb R^k$ can be written as
$$ \hat \theta^* = \theta \ \mathbb E(\mathbf T^\top) \big[\mathbb E (\mathbf T \mathbf T^\top ) \big]^{-1} \mathbf T,  $$ 
where the term $\mathbb E(\mathbf T^\top) \big[\mathbb E (\mathbf T \mathbf T^\top ) \big]^{-1} \mathbf T$ appears as an inadequate estimate of $1$. One can argue that aiming for the oracle in this case may divert from the primary objective to estimate $\theta$. \\

Besides the reasons to rule out linear averaging in this framework, one can provide some additional arguments in favor of the affine constraint $\lambda^\top \operatorname{J} = \operatorname{I}$ as a minimal requirement on $\lambda$. For instance, remark that if $\hat \lambda$ and $\lambda^*$ satisfy this condition, the error term can be written as 
$$ (\hat \lambda - \lambda^*)^\top  \mathbf T = (\hat \lambda - \lambda^*)^\top  (\mathbf T - \operatorname{J} \theta ),$$ 
where both  $(\hat \lambda - \lambda^*)$ and $(\mathbf T - \operatorname{J} \theta)$ can contribute to make  the error term negligible. Moreover, this restriction leads to an expression of the mean squared error matrix that only involves $\Sigma$, as pointed out in \eqref{traceexp}. Thus, building an approximation $\hat \lambda$ of the optimal combination only requires to estimate $\Sigma$ (this step is discussed in details in Section \ref{sec:mse}). Finally, the error bound proved in Theorem \ref{slutsky2multi} only holds if $\Lambda$ is a subset of $\Lambda_{\max}$ which tends to confirm the necessity of this constraint in this framework. \\

We now discuss four examples of constraint sets. The examples are given in decreasing order of the performance of the oracle, starting from the maximal constraint set $\Lambda_{\max} = \{ \lambda \in \mathbb R^{k \times d}: \lambda^\top \operatorname{J} = \operatorname{I} \}$ and ending with estimator selection. Apart from the last example, the other constraints sets are convex, thus guaranteeing a unique solution both for the oracle and the averaging estimator.

\begin{itemize}

\item When  a good estimation of $\Sigma$ can be provided, it is natural to consider the maximal constraint set $\Lambda = \Lambda_{\max}$, thus aiming for the best possible oracle. This set is actually an affine subspace of $\mathbb R^{k \times d}$ and as such, it is convex. The oracle, obtained by minimizing the convex map $  \lambda \mapsto \tr(\lambda^\top  \Sigma \lambda)$ subject to the constraint $\lambda^\top \operatorname{J} = \text I$ is given by $\hat \theta^*_{\max} = \lambda^{* \top}_{\max} \mathbf T$ where
\begin{equation}\label{oracle_max} 
\lambda^*_{\max} = \Sigma^{-1} \operatorname{J} (\operatorname{J}^\top  \Sigma^{-1} \operatorname{J})^{-1},
\end{equation}
generalizing the formula given in Section~\ref{ag_oneparameter}. Its mean squared error can be calculated directly
$$ \mathbb E \big[ (\hat \theta^*_{\max} - \theta) (\hat \theta^*_{\max} - \theta )^\top \big]  =  (\operatorname{J}^\top  \Sigma^{-1} \operatorname{J})^{-1}. $$
One verifies that $\lambda^*_{\max}$ is a minimizer by
\begin{equation}\label{optim_max}   \lambda^\top  \Sigma \lambda - (\operatorname{J}^\top  \Sigma^{-1} \operatorname{J})^{-1 } =   \lambda^\top  \Sigma \lambda  - \lambda^{* \top}_{\max}  \Sigma \lambda^*_{\max}  = (\lambda - \lambda^*_{\max})^\top  \Sigma (\lambda - \lambda^*_{\max}) \end{equation}
which holds for all $\lambda \in \Lambda_{\max}$ due to the condition $\lambda^\top  \operatorname{J} = \text I$, and where the last matrix is non-negative definite.

Moreover, \eqref{optim_max} shows that the oracle is not only the solution of our optimization problem \eqref{oracle}, but it also fulfills the stronger requirement \eqref{matrixoracle}.  In particular each component $\hat \theta^*_{\max,j}$ of the oracle is the best linear transformation $\lambda^\top\mathbf T$,  $\lambda\in\Lambda_{\max}$, that one can get to estimate $\theta_{j}$. Another desirable property of the choice $\Lambda=\Lambda_{\max}$ is that due to the closed expression \eqref{oracle_max},  the averaging estimator  $\hat \theta_{\max}$ obtained by replacing $\Sigma$ by its estimation $\hat \Sigma$ has also a closed expression which makes it easily computable, namely
\begin{equation}\label{ag_max} 
 \hat \theta_{\max} =  (\operatorname{J}^\top  \hat \Sigma^{-1} \operatorname{J})^{-1} \operatorname{J}^\top \hat \Sigma^{-1} \mathbf T.
 \end{equation}
As mentioned earlier, the maximal constraint set allows one to use the information contained in external collections to estimate each parameter. This requires to estimate the whole MSE matrix, including the cross correlations between different collections $\mathbf T_i$. While this can produce surprisingly good results in some cases (see Section~\ref{simus}), it may deteriorate the estimator if the external collections do not contain significant additional information on the parameter. 

\item A simpler framework is to consider component-wise averaging, for which only the collection $\mathbf T_j$ is involved in the estimation of $\theta_{j}$. The associated set of weights is the set of matrices $\lambda$ whose support is included in the support of $\operatorname{J}$, that is
$$ \Lambda = \{ \lambda \in \Lambda_{\max}: \ \text{supp}(\lambda) \subseteq \text{supp}(\operatorname{J}) \},$$ 
where for a matrix $A=(A_{i,j})\in\mathbb R^{k\times d}$, $\text{supp}(A):=\{(i,j),\, A_{i,j}\neq0\}$.  
In this particular framework, the covariance of two initial estimators in different collection $\mathbf T_i, \mathbf T_j$, $i \neq j$ is not involved in the computation of the oracle, so that the corresponding entries of $\Sigma$ need not be estimated. Consequently, each component of $\theta$ is combined regardless of the others and as a result, the oracle is given by 
$$ \hat \theta^*_j = \frac{ \mathbf 1^\top  \Sigma_j^{-1} \mathbf T_j}{ \mathbf 1^\top  \Sigma_j^{-1} \mathbf 1 }, \ j=1,...,d, $$
where 
$$\Sigma_j = \mathbb E \big[ (\mathbf T_j - \theta_{j} \mathbf 1)( \mathbf T_j - \theta_{j} \mathbf 1)^\top \big] \in \mathbb R^{k_j \times k_j},\ j =1,...,d. $$ 
In order to build the averaging estimator, it is sufficient to plug an estimate of $\Sigma_j$ for $j =1,...,d$ in the above expression, which makes it easily computable. See Section \ref{sec5.2} for further discussion.

\item Convex averaging  corresponds  to the choice
\begin{equation}\label{convex_weights} \Lambda = \{ \lambda \in \Lambda_{\max}: \ \lambda_{i,j} \geq 0, \ i=1,...,k, \ j=1,...,d \}.\end{equation}
 Observe that the positivity restriction combined with the condition $\lambda^\top  \operatorname{J} = \text I$ results in $\lambda$ having its support included in that of $\operatorname{J}$, making convex averaging a particular case of component-wise averaging. This means that each component of $\theta$ can be dealt with separately. So, for sake of simplicity in this example, we only consider the case $d=1$. 

Convex combination of estimators is a natural choice that has been widely studied in the literature. An advantage lies in the increased stability of the solution, due to the restriction of $\lambda$ to a compact set, though the oracle may of course be less efficient than in the maximal case $\Lambda=\Lambda_{\max}$. The use of convex combinations is also particularly convenient to preserve some properties of the initial estimators, such as positivity or boundedness. Moreover, imposing non-negativity often leads to sparse solutions. 

In this convex constrained optimization problem, the minimizer $\hat \lambda = \arg \min_{\lambda \in \Lambda} \ \lambda^\top  \hat \Sigma \lambda$ can either lie in the interior of the domain, in which case $\hat  \lambda =  \hat \Sigma^{-1} \mathbf 1 /\mathbf 1^\top  \hat \Sigma^{-1} \mathbf 1$ corresponds to the global minimizer over $\Lambda_{\max}$, or on the edge, meaning that it has at least one zero coordinate. Letting $\hat m \subseteq \{1,...,k \}$ denote the support of $\hat \lambda$, it follows that the averaging procedure obtained with the estimators $\mathbf T_{\hat m} :=( T_i)_{ i \in \hat m}$ leads to a solution $\hat \lambda_{\hat m}$ with full support. As a result, it can be expressed as the global minimizer for the collection $\mathbf T_{\hat m}$,
$$ \hat \lambda_{\hat m} =  \frac{\hat \Sigma^{-1}_{\hat m} \mathbf 1 }{\mathbf 1^\top  \hat \Sigma^{-1}_{\hat m} \mathbf 1},  $$
where $\hat \Sigma_{\hat m}$ is the submatrix composed of the entries $\hat \Sigma_{i,j}$ for $(i,j) \in \hat m^2$. Since we have by construction $\hat \lambda_{\hat m}^\top  \mathbf T_{\hat m} = \hat \lambda^\top  \mathbf T = \hat \theta$, we deduce the following characterization of the convex averaging solution:
$$ \hat \theta =  \frac{ \mathbf 1 ^\top \hat \Sigma^{-1}_{\hat m} \mathbf T_{\hat m}}{\mathbf 1^\top  \hat \Sigma^{-1}_{\hat m} \mathbf 1},  $$
where $\hat m$ is the admissible support with minimal mean squared error, i.e. $ \hat m = \arg \max_{m \subseteq \{1,...,k \}} \ \mathbf 1 ^\top  \hat \Sigma^{-1}_m \mathbf 1$ subject to the constraint that $ \hat \Sigma_m^{-1} \mathbf 1$ has all its coordinates positive. This provides an easy method to implement convex averaging in practice. Remark that this method is only efficient if $k$ is not too large, otherwise we recommend to use a standard quadratic programming solver to get $\hat \lambda$, see for instance \cite{nocedal2006}.

\item The last example deals with estimator selection viewed as a particular case of averaging. Performing estimator selection based on an estimation of the MSE is an approach used in numerous practical situations. In the univariate case, estimator selection corresponds to the constraint set
$$ \Lambda = \big\{ (1,0,...,0)^\top , (0,1,0,...,0)^\top ,...,(0,...,0,1)^\top \big\}.   $$
The main advantage of this framework is that it only requires to estimate the mean squared error of each estimator $T_j$, i.e., the diagonal entries of $\Sigma$. Applying the procedure in this case simply consists in selecting the $T_j$ with minimal estimated mean squared error. 

While the oracle is easier to approach  in this framework, estimator selection may suffer from the poor efficiency of the oracle, compared to the previous examples. For this reason, we do not recommend to settle for estimator selection if one can provide a reasonable estimation of $\Sigma$, as the oracle under larger constraint sets can be much more efficient while remaining reachable. This observation is confirmed by the numerical study of Section \ref{simus} where the average estimator appears to be better than the best estimator in the initial collection in most cases. Finally, remark that the theoretical performance of estimator selection is not covered by Theorem \ref{slutsky2multi}, due to the non-convexity of the constraint set.

\end{itemize}

\subsection{Estimation of the MSE matrix}\label{sec:mse}

The accuracy of $\hat \Sigma$ is clearly a main factor to the performance of the averaging method. There exist several methods to construct $\hat \Sigma$, whether the model is parametric or not. In all cases, the estimation of $\Sigma$ can be carried out from the same data as those used to produce the initial estimators $T_i$ and no sample splitting is needed. \\

In a fully specified parametric model, the MSE matrix $\Sigma$ can be estimated by plugging an initial estimate of $\theta$. Precisely, assuming that the MSE matrix can be expressed as the image of $\theta$ through a known continuous map $\Sigma(.): \mathbb R^d \to \mathbb R^{k \times k}$, one can choose $\hat \Sigma = \Sigma(\hat \theta_0)$, where $\hat\theta_0$ is a consistent estimate of $\theta$. A suitable choice for $\hat\theta_0$ is to take one of the initial estimators if it is known to be consistent, or the average $\frac 1 k \sum_{i=1}^k T_i$ provided all initial estimators are consistent. If the map $\Sigma(.)$ is not explicitly known, $\Sigma(\hat\theta_0)$ may be approximated by Monte-Carlo simulations of the model using the estimated parameter $\hat \theta_0$, a procedure sometimes called parametric bootstrap. This method is illustrated in our examples in Sections~\ref{sec5.2} and \ref{sec:bool} and reveals to be efficient whenever the model is well specified. Remark that in this parametric situation, the averaging procedure does not require any information other than the initial collection $\mathbf T$.

In some cases,  $\Sigma$ may also depend on a nuisance parameter $\eta$. In this situation, $\hat \Sigma$ can be built similarly by plugging or Monte-Carlo, provided $\eta$ can be estimated from the observations. This situation requires the sample $X_1,...,X_n$ used to built the initial estimators $T_i$ to be available to the user.\\

In a semi and non-parametric setting, a parametric closed-form expression for $\Sigma$ may be available asymptotically, i.e. when the sample size on which $\mathbf T$ is built tends to infinity, and the above plugging method then becomes possible, see also (i)-(iii) in Section~\ref{sec:asymptotic}. Alternatively, $\Sigma$ can be estimated by standard bootstrap if no extra information is available. These two methods are implemented in the first example of Section~\ref{simus}.

\section{Theoretical results}\label{theory}

\subsection{Non-asymptotic error bound}\label{errorbound}

The performance of the averaging estimator relies on the accuracy of $\hat \Sigma$, but more specifically, on the ability to evaluate tr$(\lambda^\top  \Sigma \lambda)$ as $\lambda$ ranges over $\Lambda$. As a result, it is not crucial that $\hat \Sigma$ be a perfect estimate of $\Sigma$ as long as the error $\vert \tr(\lambda^\top  \hat \Sigma \lambda) - \tr(\lambda^\top \Sigma \lambda) \vert$ is small for $\lambda \in \Lambda$, which can hopefully be achieved by a suitable choice of the constraint set. In order to measure the accuracy of $\hat \Sigma$ for this particular purpose, we introduce the following criterion. For two symmetric positive definite matrices $A$ and $B$ and for any  non-empty set $\Lambda$ that does not contain $0$, let $ \delta_\Lambda(A \vert B)$ denote the maximal divergence of the ratio $\tr ( \lambda^\top  A \lambda)/\tr ( \lambda^\top  B \lambda)$ over $\Lambda$,
$$ \delta_\Lambda(A \vert B)  = \sup_{\lambda \in \Lambda} \ \left| 1 - \frac{ \tr ( \lambda^\top  A \lambda)}{\tr ( \lambda^\top  B \lambda)} \right|,  $$
and $ \delta_\Lambda(A , B) = \max \{ \delta_\Lambda(A \vert B), \delta_\Lambda(B \vert A) \}$. We are now in position to state our main result.

\begin{theorem}\label{slutsky2multi} Let $\Lambda$ be a non-empty closed convex subset of $\Lambda_{\max}$ with associated oracle $\hat \theta^*$ defined through \eqref{oracle}, and $\hat \Sigma$ a symmetric positive definite $k \times k$ matrix. The averaging estimator $\hat \theta = \hat \lambda^\top  \mathbf T$ defined through \eqref{hattheta} satisfies
\begin{equation}\label{ineg_oracle}
 \Vert \hat \theta - \hat \theta^*\Vert ^2 \leq  \tilde\delta_\Lambda (\hat \Sigma^{}, \Sigma^{} )  \   \Vert \mathbf S \Vert^2  \  \mathbb E \Vert \hat \theta^* - \theta \Vert^2  \end{equation}
  where $\tilde\delta_\Lambda (\hat \Sigma^{}, \Sigma^{} ) = 2 \delta_\Lambda(\hat \Sigma^{}, \Sigma^{}) + \delta_\Lambda (\hat \Sigma^{}, \Sigma^{} )^2$ and  $\mathbf S= \Sigma^{- \frac 1 2} (\mathbf T - \operatorname{J} \theta)$.
\end{theorem}

In this theorem, we provide an upper bound on the distance of the averaging estimator to the oracle. We emphasize that this result holds without requiring any condition on the joint behavior of $\mathbf T$ and $\hat \Sigma$ (in particular, they may be strongly dependent). Moreover, we point out that the upper bound applies to the actual error to the oracle (for the current event $\omega$), contrary to classical oracle inequalities which generally involve an expected loss of some kind. Nonetheless,  the following corollary compares the mean squared errors of the averaging estimator and the oracle.

\begin{corollary}\label{cortheoric} Under the assumptions of Theorem \ref{slutsky2multi}, 
for all $\epsilon>0$, 
\begin{equation}
\mathbb E \Vert \hat \theta - \theta\Vert ^2 \leq (1+\epsilon) \ \mathbb E \Vert \hat \theta^* - \theta \Vert^2 
\left[1+\epsilon^{-1}  \mathbb E \big(\tilde\delta_\Lambda (\hat \Sigma^{}, \Sigma^{} )  \ \Vert \mathbf S \Vert^2 \big)\right].
\end{equation}
\end{corollary}
Some  comments on Theorem~\ref{slutsky2multi} and its corollary are in order.

\begin{itemize}
	\item Recall that a main concern when selecting the constraint set $\Lambda$ is to be able to make the distance to the oracle $\Vert \hat \theta - \hat \theta^* \Vert$ negligible compared to the error of the oracle $\Vert \hat \theta^* - \theta \Vert$. This objective is achieved  whenever the  term  $\tilde\delta_\Lambda (\hat \Sigma^{}, \Sigma^{} )  \Vert \mathbf S \Vert^2$  in \eqref{ineg_oracle} can be shown to be negligible. Corollary~\ref{cortheoric} makes this remark more specific in the $\mathbb L^2$ sense: by choosing $\epsilon$ tending to 0 not too fast, the mean squared errors of the averaging estimator and the oracle are seen to be asymptotically equivalent whenever $\mathbb E \big(\,\tilde\delta_\Lambda (\hat \Sigma^{}, \Sigma^{} )   \Vert \mathbf S \Vert^2 \big)$ tends to 0. Section~\ref{sec:asymptotic} details more consequences of Theorem~\ref{slutsky2multi} from an asymptotic point of view.

	\item The  factor $\tilde\delta_\Lambda (\hat \Sigma^{}, \Sigma^{} )$, which only involves the divergence $\delta_\Lambda (\hat \Sigma^{}, \Sigma^{})$, emphasizes the influence of the constraint set $\Lambda$ and the accuracy of $\hat\Sigma$ to estimate $\Sigma$. It appears that while the efficiency of the oracle is increased for large sets $\Lambda$, one must settle for combinations $\lambda$ for which $\tr(\lambda^\top \Sigma \lambda)$ can be well evaluated, in order to get a small value of $\delta_\Lambda (\hat \Sigma^{}, \Sigma^{})$. Actually, as stated in  Lemma \ref{lem} of the Appendix, we have
\begin{equation}\label{Ldelta}  \delta_\Lambda (\hat \Sigma^{}, \Sigma^{}) \leq  \Vert \hspace{-0.04cm} \vert \hat \Sigma^{} \Sigma^{- 1}  - \Sigma^{} \hat \Sigma^{- 1}  \Vert \hspace{-0.04cm} \vert, \end{equation}
where $\Vert \hspace{-0.04cm} \vert . \Vert \hspace{-0.04cm} \vert$ denotes the operator norm. This shows that the efficiency of the averaging procedure can be measured by how well $\hat \Sigma^{} \Sigma^{- 1}$ approximates the identity matrix. 

It is difficult to study the behavior of $\delta_\Lambda (\hat \Sigma^{}, \Sigma^{})$ or even $\hat \Sigma^{} \Sigma^{- 1}$ without more information on the statistical model from which $\mathbf T$ is computed. In particular, we are not able to derive oracle-like inequalities 
involving minimax rates of convergence without further specification. For instance, $\delta_\Lambda (\hat \Sigma^{}, \Sigma^{})$ can be expected to converge in probability to zero at a typical rate of $\sqrt n$ in a well specified parametric sampling model, while similar properties should be seldom verified in semi or non-parametric models. 

	\item Finally, the term $ \Vert \mathbf S \Vert^2$ in \eqref{ineg_oracle}  shows the price to pay for averaging too many estimators, in view of the equality
$$ \mathbb E \Vert \mathbf S \Vert^2=\mathbb E \Vert \Sigma^{- \frac 1 2} (\mathbf T - \operatorname{J} \theta) \Vert^2 =  k. $$	
This suggests that the averaging procedure might be improved by performing a preliminary selection to keep only the relevant estimators. Moreover, including too many estimators to the initial collection increases the possibilities of strong correlations, which may lead to a near singular matrix $\Sigma$ and result in amplified errors when computing $\hat \Sigma^{-1}$ and a larger value of $\delta_\Lambda (\hat \Sigma^{}, \Sigma)$.
\end{itemize}

\subsection{Asymptotic study}\label{sec:asymptotic}

The properties of the averaging estimator established in Theorem \ref{slutsky2multi} do not rely on any assumption on the construction of $\mathbf T$ or $\hat \Sigma$. In this section, we investigate the asymptotic properties of the averaging estimator in a situation where both $\mathbf T$ and $\hat \Sigma$ are computed from a set of observations $X_1,\dots,X_n$ of size $n$ growing to infinity. From now on, we modify our notations to $\mathbf T_n$, $\hat \Sigma^{}_{n}$, $\Sigma^{}_{n}$, $\lambda^*_n$, $\hat \lambda_n$, $\hat \theta_n$ and $\hat \theta^*_n$ to emphasize the dependency on $n$. \bigskip

In practice, we expect the oracle $\hat \theta^*_n$ to satisfy good properties such as consistency and asymptotic normality. Theorem~\ref{slutsky2multi} suggests that $\hat \theta_n$ should inherit these asymptotic properties if $\Sigma_n$ can be sufficiently well estimated. Remark that if the initial estimators $T_i$ are consistent in quadratic mean, $\Sigma_n$ converges to the null matrix as $n \to \infty$. In this case, providing an estimator $\hat \Sigma_n$ such that $\hat \Sigma_n - \Sigma_n  \overset{p}{\longrightarrow} 0 $ is clearly not sufficient for $\hat \theta_n$ to achieve the asymptotic performance of the oracle (here $p$ stands for the convergence in probability while $d$ is used for distribution). On the contrary, requiring that $\hat \Sigma_n^{-1} - \Sigma_n^{-1} \overset{p}{\longrightarrow} 0$ is unnecessarily too strong and would be nearly impossible to achieve. In fact, we show in Proposition \ref{cor} below that the condition
\begin{equation}\label{hypasymptotic} \hat \Sigma^{}_{n} \Sigma^{-1}_{n}\overset{p}{\longrightarrow} \mathrm{I}\end{equation} 
appears as a simple compromise, both sufficient for asymptotic optimality and reasonable enough to be verified in numerous situations with regular estimators of $\Sigma_n$. We briefly discuss a few examples.

\begin{itemize}
	\item[(i)] If $\sqrt n (\mathbf T_n - \operatorname{J} \theta)$ converges in $\mathbb L^2$ to a Gaussian vector $\mathcal N(0,W)$ with $W$ a non-singular matrix, providing a consistent estimator, say $\hat W_n$, of $W$ is sufficient to verify \eqref{hypasymptotic}, taking $\hat \Sigma_n = n^{-1}\hat W_n$. The situation becomes particularly convenient if the limit matrix $W$ follows a known parametric expression $W = W(\eta,\theta)$, with $\eta$ a nuisance parameter (see the first example in Section~\ref{simus}). If the map $W(.,.)$ is continuous, plugging consistent estimators $\hat \eta_0$, $\hat \theta_0$ yields an estimator $\hat W^{}_{n} = W(\hat \eta_0, \hat \theta_0)$ that fulfills \eqref{hypasymptotic}.  If the map $W(.,.)$ is unknown, parametric bootstrap  based on the initial estimates $\hat \eta_0$ and $\hat \theta_0$ may be used and the same theoretical justifications apply. 
Observe that knowing the rate $\sqrt n$ in this example is not necessary as it simplifies in the expression of $\hat \theta_n$. In fact, a different rate of convergence, even unknown, would lead to the exact same result. In this case, the asymptotic normality can make it possible to construct asymptotic confidence intervals of minimal length for the parameter, as shown in Proposition \ref{cor} below.

	\item[(ii)] More generally, if $\Sigma_n$ satisfies 
\begin{equation}\label{meme_vitesse} \Sigma^{}_{n} = a_n W + o(a_n), \end{equation}
for some vanishing sequence $a_n$, building a consistent estimator of $W$ is sufficient to achieve \eqref{hypasymptotic}. Here again, the rate of convergence needs not be known. 
	
	\item[(iii)] If we have different rates of convergence within the collection $\mathbf T_n$, the condition \eqref{hypasymptotic} can be verified if the normed eigenvectors of $\Sigma^{}_{n}$ converge as $n \to \infty$. Precisely, if there exist an orthogonal matrix $P$ (i.e. with $P^\top  P = \operatorname I$) and a known deterministic sequence $(A_n)_{n \in \mathbb N}$ of diagonal invertible matrices such that 
$$ \lim_{n \to \infty} \ A_n P \Sigma^{}_{n} P^\top  = D,$$
for some non-singular diagonal matrix $D$, producing consistent estimators $\hat P_n$ and $\hat D_n$ of $P$ and $D$ respectively enables to verify \eqref{hypasymptotic} by $\hat \Sigma^{}_{n} = \hat P_n^\top  A_n^{-1} \hat D_n \hat P_n$. Here, the limit of the normed eigenvectors of $\Sigma^{}_{n}$ are given by the rows of $P$ and the estimator $\hat \Sigma^{}_{n}$ is constructed from the asymptotic expansion of $\Sigma^{}_{n}$. This example allows to have different rates of convergence within the collection $\mathbf T_n$ but also covers the previously mentioned examples where all constant combinations $\lambda^\top  \mathbf T_n $ converge to $\theta$ at the same rate.
	
\end{itemize}

Let us introduce some additional definitions and notation. For each component $\theta_j$, $j=1,...,d$, we define
$$ \alpha_{n,j} :=  \mathbb  E \Vert \hat \theta^*_{n,j} - \theta_j \Vert^2 = \underline{\lambda}_{n,j}^{* \top} \ \Sigma_{n} \ \underline{\lambda}_{n,j}^*, $$
where we recall that $\underline{\lambda}_{n,j}^*$ is the $j$-th column of $\lambda^*_n$. Similarly, let $\hat \alpha_{n,j} = \underline{\hat \lambda}_{n,j}^{\top} \hat \Sigma_{n} \underline{\hat \lambda}_{n,j}$. We assume that the quadratic error of the oracle, given by
$$ \alpha_n := \mathbb  E \Vert \hat \theta^*_n - \theta \Vert^2 = \tr(\lambda_n^{* \top}  \Sigma^{}_n \lambda^*_n) = \sum_{j=1}^d \alpha_{n,j}, $$
converges to zero as $n \to \infty$. For a given constraint set $\Lambda\subset\mathbb R^{k\times d}$, we denote by $ \Lambda_j = \{ \underline{\lambda}_{j}: \lambda \in \Lambda \} \subset \mathbb R^k$ its marginal set. We say that $\Lambda$ is a \textit{cylinder} if $\Lambda =\{\lambda: \underline{\lambda}_{1} \in \Lambda_1,..., \underline{\lambda}_d \in \Lambda_d \}$, i.e., if $\Lambda$ is the Cartesian product of its marginal sets $\Lambda_j$. We point out that choosing a constraint set $\Lambda$ that satisfies this property is not restrictive in general, as it simply states that each  vector of weights $\underline{\lambda}_j$ used to estimate $\theta_j$ can be computed independently of the others.  In particular, all the constraint sets discussed in Section \ref{sec:set} are cylinders. 

\begin{proposition}\label{cor} If \eqref{hypasymptotic} holds, then
\begin{equation}\label{weakasopt} \Vert \hat \theta_n - \theta \Vert^2 = \Vert \hat \theta^*_n - \theta\Vert ^2 + o_p(\alpha_n). \end{equation}
Moreover, if $\Lambda$ is a cylinder and $\alpha_{n,j}^{-\frac 1 2} (\hat \theta^*_{n,j} - \theta_j)  \overset{d}{\longrightarrow} \mathcal Z$  for some $j=1,...,d$, where $\mathcal Z$ is a random variable, then
\begin{equation}\label{lawasopt}\hat \alpha_{n,j}^{-\frac 1 2}  (\hat \theta_{n,j} - \theta_j) \overset{d}{\longrightarrow} \mathcal Z. \end{equation}
\end{proposition}

This proposition establishes that building an estimate $\hat \Sigma_n$ for which \eqref{hypasymptotic} holds ensures that the 
error of the average $\hat \theta_n$ is asymptotically comparable to that of the oracle, up to $o_p(\alpha_n)$.  In addition, under the mild assumption that $\Lambda$ is a cylinder, it is possible to provide an asymptotic confidence interval for each $\theta_j$ when the limit distribution $\mathcal Z$ is known. In most applications,  this distribution is a standard Gaussian, as for instance under the assumptions in (i), but other more complicated situations may be handled as discussed in the following remark.  From \eqref{hattheta}, we know these confidence intervals are of minimal length amongst all possible confidence intervals based on a linear combination of $\mathbf T_{n}$. Note that no extra estimation is needed to compute $\hat \alpha_{n,j}$, as it is entirely determined by $\hat\lambda_n$ and $\hat\Sigma_n$.

\begin{rem} \label{remarque}
As noticed above, $\mathcal Z$ can be expected to follow a standard Gaussian distribution in numerous situations. For instance, this happens under the setting of (i), with a possibly different normalization than $\sqrt n$, or under the setting of (iii) provided $\mathbf T_n$ is asymptotically normal with vanishing bias. We refer to Section~\ref{simus} for actual examples. However, in presence of misspecified initial estimators, the distribution of $\mathcal Z$ can be different. In \cite{hjort2003frequentist}, the authors study a specific local misspecification framework where the asymptotic law of the oracle is not Gaussian, neither centered (see their Theorem 4.1 and Section 5.4 for an averaging procedure based on the mean squared error). In these cases, the quantiles of $\mathcal Z$ may depend on some unknown extra parameters that have to be estimated to provide asymptotic confidence intervals. In such a misspecified framework though, the challenge is to estimate accurately $\Sigma$. Conditions implying \eqref{hypasymptotic} in a misspecified framework are difficult to establish and are beyond the scope of the present paper.
\end{rem}

In Proposition~\ref{cor}, the asymptotic optimality of $\hat \theta_n$ is stated in probability. In view of Corollary~\ref{cortheoric}, it is not difficult to strengthen this result to get the asymptotic optimality in quadratic loss, i.e.
\begin{equation}\label{l2opt} \mathbb E \Vert \hat \theta_n - \theta \Vert^2 = \mathbb E \Vert \hat \theta^*_n - \theta \Vert^2 (1 + o(1)), \end{equation}
provided additional assumptions hold. If for instance $\hat \Sigma_n$ and $\mathbf T_n$ are computed from independent samples (which may be achieved by sample splitting), then \eqref{l2opt} holds as soon as $\mathbb E [\tilde\delta_\Lambda(\hat \Sigma_n, \Sigma_n)] $ tends to $ 0$, which is achieved if $\hat \Sigma^{}_{n} \Sigma^{-1}_{n}$ and $\Sigma^{}_{n} \hat \Sigma^{-1}_{n}$ tend to the identity matrix in $\mathbb L^2$. We emphasize however that the use of sample splitting may reduce the performance of the oracle, as it would be computed from fewer data. One can argue that this is a high price to pay to obtain asymptotic optimality in $\mathbb L^2$ and is not to be recommended in this framework.
Asymptotic optimality in $\mathbb L^2$ can also be achieved if one can show there exists $p > 1$ such that
 $$ \sup_{n \in \mathbb N} \ \mathbb E \Vert \Sigma_n^{- \frac 1 2} (\mathbf T_n - \operatorname{J} \theta) \Vert^\frac{2p}{p-1} < \infty \quad \text{and} \quad \lim_{n \to \infty} \ \mathbb E [ \tilde\delta_\Lambda (\hat \Sigma^{}_{n}, \Sigma^{}_{n} )^{p}] = 0,$$
which is a direct consequence of Corollary~\ref{cortheoric}  by applying H\"older's inequality. These conditions ensure the asymptotic optimality in $\mathbb L^2$ of the averaging estimator without sample splitting, but they remain nonetheless extremely difficult to check in practice.

\section{Applications}\label{simus}

This section gathers four examples of models where we apply our averaging procedure. Depending on the situation, we combine parametric, semi-parametric or non-parametric estimators. The examples in  Section~\ref{sec5.2} and \ref{sec:bool} involve a bivariate parameter which allows us to assess the multivariate procedure introduced in  Section~\ref{sec:agseveral}. For the estimation of the MSE matrix $\Sigma_n$, we use either parametric bootstrap, (non-parametric) bootstrap or the asymptotic expression $\Sigma_\infty$ when it has a parametric form. From a theoretical point of view, all our examples satisfy the main  condition \eqref{hypasymptotic}  implying the asymptotic optimality of the average estimator in the  sense of Proposition~\ref{cor}, since they fit either (i) or (ii) in Section~\ref{sec:asymptotic}, except when non-parametric bootstrap or misspecified models are used. The last two settings are common situations, so we chose to include them in our simulation study, however they are out of the scope of the theoretical study of this paper and will be the subject to future investigations.

\subsection{Estimating the position of a symmetric distribution}\label{mean_med}
Let us consider a continuous  real distribution with density $f$,  symmetric around some parameter $\theta$.
To estimate $\theta$ from a sample  of $n$ realisations $x_1,\dots,x_n$, simple solutions are to use  the mean $\overline x_n$ or the median $x_{(n/2)}$. Both estimators are consistent whenever  $\sigma^2=\int (x-\theta)^2 f(x) dx$ is finite.\\

As noticed in Section~\ref{intro},  the idea of combining the mean and the median to construct a better estimator goes back to Pierre Simon de Laplace \cite{laplace}. P. S. de Laplace obtains the expression of the weights in $\Lambda_{\max}$ that ensure a minimal asymptotic variance for the averaging estimator. In particular, he deduced  that for a Gaussian distribution, the better combination is to take the mean only, showing for the first time the efficiency of the latter. For other distributions, he noticed that  the best combination is not available in practice because it depends on the unknown distribution.

Similarly, we consider the averaging of the mean and the median over $\Lambda_{\max}$. We have two initial estimators  $T_1=\overline x_n$, $T_2=x_{(n/2)}$ and the averaging estimator  is given by \eqref{ag_max} where $\operatorname{J}$ is just in this case the vector $(1,1)^\top $. We assume that the $n$ realisations are independent and we propose two ways to estimate the MSE matrix $\Sigma_n$:
\begin{enumerate}
\item {\it Based on the asymptotic equivalent of $\Sigma_n$}. 
 The latter, obtained in P. S. de Laplace's work and recalled in \cite{stiegler}, is $n^{-1}W$ where
$$W=\begin{pmatrix}
\sigma^2 & \frac{\E |X-\theta|}{2f(\theta)} \\
\frac{\E |X-\theta|}{2f(\theta)} & \frac{1}{4 f(\theta)^2}
\end{pmatrix}.$$
Each entry of $W$ may be estimated from an initial consistent estimate $\hat\theta_0$ of $\theta$ as follows:  $\sigma^2$ by the empirical variance $s_n^2$; $\E |X-\theta|$ by $\hat m=1/n\sum_{i=1}^n |x_i-\hat\theta_0 |$; and $f(\theta)$ by the kernel estimator $\hat f(\hat\theta_0)=1/(nh) \sum_{i=1}^n \exp(-(x_i-\hat\theta_0)^2/(2h^2) )$, where $h$ is chosen, e.g.,  by the so-called Silverman's rule of thumb  (see \cite{silverman}). With this estimation of $\Sigma_n$, we get the following averaging estimator:
\begin{equation}\label{theta_AG}
\hat\theta_{AV}=\frac{p_1}{p_1+p_2} \overline x_n +  \frac{p_2}{p_1+p_2} x_{(n/2)}
\end{equation}
where $p_1=1/(4\hat f(\hat\theta_0)) - \hat m /2$ and $p_2=s_n^2 \hat f(\hat\theta_0)- \hat m/2$.  This  estimator corresponds to an empirical version of the best combination obtained by  P. S. de Laplace.

\item {\it Based on bootstrap}. We draw  with replacement $B$ samples of size $n$ from the original dataset. We compute the mean and the median of each sample, respectively denoted $\overline x^{(b)}_n$ and  $x_{(n/2)}^{(b)}$ for $b=1,\dots,B$. The MSE matrix $\Sigma_n$ is then estimated by 
$$\frac 1 B \begin{pmatrix} \sum_{b=1}^B (\overline x^{(b)}_n-\overline x_n)^2 &  \sum_{b=1}^B (\overline x^{(b)}_n-\overline x_n)(x_{(n/2)}^{(b)}-x_{(n/2)}) \\ \sum_{b=1}^B (\overline x^{(b)}_n-\overline x_n)(x_{(n/2)}^{(b)}-x_{(n/2)})  & \sum_{b=1}^B
(x_{(n/2)}^{(b)}-x_{(n/2)}) ^2 \end{pmatrix}.$$
This leads to another averaging estimator, denoted  by $\hat\theta_{AVB}$.

\end{enumerate}
Let us note that the first procedure above fits the asymptotic justification presented in example (i) of Section~\ref{sec:asymptotic}. For this reason, $\hat\theta_{AV}$ is asymptotically as efficient as the oracle, provided $\hat\theta_0$ is consistent. Moreover, the oracle is asymptotically Gaussian and  an asymptotic confidence interval for $\theta$ can be provided without further estimation, see Section~\ref{sec:asymptotic} and particularly Remark~\ref{remarque}.  For the second procedure, theory is lacking to study the behaviour of $\tilde\delta_\Lambda$ in \eqref{ineg_oracle} when $\hat\Sigma_n$ is estimated by  bootstrap, so no consistency can be claimed at this point. \\ 

Table~\ref{tab:mean_med} summarizes the estimated MSE of $\overline x_n$, $x_{(n/2)}$, $\hat\theta_{AV}$ and $\hat\theta_{AVB}$,  for  $n=30,\, 50,\, 100$,  and for different distributions, namely: Cauchy, Student with 5 degrees of freedom, Student with 7 degrees of freedom, Logistic, standard Gaussian, and an equal mixture distribution of a $\mathcal N(-2,1)$ and a $\mathcal N(2,1)$. For all distributions, $\theta=0$.
For the initial estimate $\hat\theta_0$ in \eqref{theta_AG}, we take the median  $x_{(n/2)}$, because it is  well defined and consistent  for any continuous distribution. The number of bootstrap samples taken for $\hat\theta_{AVB}$ is $B=1000$.

\begin{table}[htbp]
\resizebox{\textwidth}{!} {
\begin{tabular}{lcccccccccccc}
\hline
 & \multicolumn{ 4}{c}{$n=30$} & \multicolumn{ 4}{c}{$n=50$} & \multicolumn{ 4}{c}{$n=100$} \\ 
  & MEAN & MED & AV & AVB \hspace{0.2cm} & \hspace{0.2cm} MEAN & MED & AV & AVB \hspace{0.2cm} & \hspace{0.2cm} MEAN & MED & AV & AVB \\ \hline
{\footnotesize Cauchy} & 2.$10^6$ & 9 & 8.95 & 8.99 \hspace{0.2cm} & \hspace{0.2cm} 4.$10^7$ & 5.07 & 4.92 & 4.9 \hspace{0.2cm} & \hspace{0.2cm} 2.$10^7$ & 2.56 & 2.49 & 2.49 \\ 
 & (1.$10^6$) & (0.14) & (0.15) & (0.15) \hspace{0.2cm} & \hspace{0.2cm} (4.$10^7$) & (0.08) & (0.08) & (0.08)  \hspace{0.2cm} & \hspace{0.2cm} (2.$10^7$) & (0.04) & (0.04) & (0.04) \\ \hline
{\footnotesize St(4)} & 6.68 & 5.71 & 5.4 & 5.43 \hspace{0.2cm} & \hspace{0.2cm} 4.12 & 3.53 & 3.33 & 3.34 \hspace{0.2cm} & \hspace{0.2cm} 1.99 & 1.74 & 1.61 & 1.62 \\ 
 & (0.1) & (0.08) & (0.08) & (0.08) \hspace{0.2cm} & \hspace{0.2cm} (0.06) & (0.05) & (0.05) & (0.05) \hspace{0.2cm} & \hspace{0.2cm} (0.03) & (0.02) & (0.02) & (0.02) \\ \hline
{\footnotesize St(7)} & 4.8 & 5.51 & 4.6 & 4.64 \hspace{0.2cm} & \hspace{0.2cm} 2.82 & 3.32 & 2.74 & 2.8 \hspace{0.2cm} & \hspace{0.2cm} 1.42 & 1.67 & 1.37 & 1.38 \\ 
 & (0.07) & (0.08) & (0.07) & (0.07) \hspace{0.2cm} & \hspace{0.2cm} (0.04) & (0.05) & (0.04) & (0.04) \hspace{0.2cm} & \hspace{0.2cm} (0.02) & (0.02) & (0.02) & (0.02) \\ \hline
{\footnotesize Logistic} & 10.89 & 12.7 & 10.76 & 10.87 & 6.64 & 7.93 & 6.52 & 6.6 & 3.3 & 4 & 3.2 & 3.26 \\ 
 & (0.16) & (0.18) & (0.16) & (0.16) & (0.09) & (0.11) & (0.09) & (0.09) & (0.05) & (0.06) & (0.05) & (0.05) \\ \hline
{\footnotesize Gauss} & 3.39 & 5.11 & 3.53 & 3.61 & 2.04 & 3.1 & 2.1 & 2.15 & 1 & 1.51 & 1.02 & 1.06 \\ 
 & (0.05) & (0.07) & (0.05) & (0.05) & (0.03) & (0.04) & (0.03) & (0.03) & (0.01) & (0.02) & (0.01) & (0.01) \\ \hline
{\footnotesize Mix} & 16.79 & 87 & 15.03 & 13.41 & 10.08 & 66.53 & 7.57 & 6.68 & 5.05 & 42.35 & 3.09 & 2.36 \\ 
 & (0.23) & (0.82) & (0.29) & (0.3) & (0.14) & (0.64) & (0.15) & (0.18) & (0.07) & (0.43) & (0.06) & (0.07) \\ \hline
\end{tabular}}
\caption{Monte Carlo estimation of the MSE of $\overline x_n$ (MEAN), $x_{(n/2)}$ (MED),  $\hat\theta_{AV}$ (AV) and  $\hat\theta_{AVB}$ (AVB) in the estimation of the position of a symmetric distribution, depending on the distribution and the sample size. The number of replications is $10^4$ and  the standard deviation of the MSE estimations is given in parenthesis.   Each entry has been multiplied by 100 for ease of presentation.}
\label{tab:mean_med}
\end{table}

While the best estimator between $\overline x_n$ and $x_{(n/2)}$ depends on the underlying distribution, the averaging estimators $\hat\theta_{AV}$ and $\hat\theta_{AVB}$ perform better than both  $\overline x_n$ and $x_{(n/2)}$, for all distributions considered in Table~\ref{tab:mean_med} except the Gaussian law. For the latter distribution, we know that the oracle is the mean, so the averaging estimator cannot improve on $\overline x_n$. However the MSE of   $\hat\theta_{AV}$ and $\hat\theta_{AVB}$ are close to that of $\overline x_n$ in this case, proving that the optimal weights $(1,0)$ are fairly well estimated.
Moreover, note that the Cauchy distribution does not belong to our theoretical setting because it has no finite moments and $\overline x_n$ should not be used.  But it turns out  that  the averaging estimators are robust in this case, as they manage to highly favor $x_{(n/2)}$. Choosing the median $x_{(n/2)}$ as the initial estimator $\hat\theta_0$  is of course crucial in this case.

Since we are in the setting of (i) in Section~\ref{sec:asymptotic}, the oracle is asymptotically normal and Proposition \ref{cor} yields an asymptotic confidence interval without any further estimation. By construction, the length of these intervals is smaller than the length of any similar confidence interval based on  $\overline x_n$ or $x_{(n/2)}$. Further, the empirical rate of coverage of these intervals is reported in Table~\ref{tab:ICmean_med} for the previous simulations, and turns out to be close to the nominal level $95\%$.

%
%
%
%

\begin{table}[htbp]
\begin{center}

\begin{tabular}{lcccccc}

 \hline
& \multicolumn{ 2}{c}{$n=30 \hspace{0.2cm}$} & \multicolumn{ 2}{c}{$n=50$} & \multicolumn{ 2}{c}{$\hspace{0.4cm} n=100$} \\ 

 & AV & AVB \hspace{0.2cm} & \hspace{0.2cm} AV & AVB \hspace{0.2cm} & \hspace{0.2cm} AV & AVB \\ \hline
\multirow{1}{*}{Cauchy} & \multicolumn{ 1}{c}{98.08} & 96.18 \hspace{0.2cm} & \hspace{0.2cm} 98.21 & 95.55 \hspace{0.2cm} & \hspace{0.2cm} 97.75 & 95.22 \\ 
\hline
\multirow{1}{*}{St(4)} & \multicolumn{ 1}{c}{93.59} & 91.45 \hspace{0.2cm} & \hspace{0.2cm} 94.38 & 92.71 \hspace{0.2cm} & \hspace{0.2cm} 94.71 & 92.55 \\ 
\hline
\multirow{1}{*}{St(7)} & \multicolumn{ 1}{c}{93.34} & 91.25 \hspace{0.2cm} & \hspace{0.2cm} 93.93 & 91.77 \hspace{0.2cm} & \hspace{0.2cm} 94.27 & 92.73 \\ 
 \hline
\multirow{1}{*}{Logistic} & \multicolumn{ 1}{c}{92.48} & 90.33 \hspace{0.2cm} & \hspace{0.2cm} 93.96 & 92.05 \hspace{0.2cm} & \hspace{0.2cm} 93.91 & 92.21 \\ 
 \hline
\multirow{1}{*}{Gauss} & \multicolumn{ 1}{c}{92.97} & 91.13 \hspace{0.2cm} & \hspace{0.2cm} 93.54 & 91.94 \hspace{0.2cm} & \hspace{0.2cm} 94.09 & 92.59 \\ 
 \hline
\multirow{1}{*}{Mix} & \multicolumn{ 1}{c}{93.19} & 93.83 \hspace{0.2cm} & \hspace{0.2cm} 94.77 & 95.97 \hspace{0.2cm} & \hspace{0.2cm} 94.94 & 97.91 \\ 
 \hline
\end{tabular}
\end{center}

\caption{Empirical rate of coverage (in $\%$) of the asymptotic $95\%$ confidence intervals based on  $\hat\theta_{AV}$ and $\hat\theta_{AVB}$ in the estimation of the position of a symmetric distribution, deduced from the same simulations as in Table~\ref{tab:mean_med}. }
\label{tab:ICmean_med}
\end{table}

Finally, while $\hat\theta_{AVB}$ suffers from a lack of theoretical justification, it behaves pretty much like $\hat\theta_{AV}$, except  for  the mixture distribution where it performs slightly better than  $\hat\theta_{AV}$.  This may be explained by the fact that $\hat\theta_{AV}$ is more sensitive than  $\hat\theta_{AVB}$ to the initial estimate $\hat\theta_0$, the variance of which is large for the mixture distribution because $f(0)$ is close to $0$. Nevertheless,  $\hat\theta_{AV}$ demonstrates very good performance in this case, for the sample sizes considered in Tables~\ref{tab:mean_med} and \ref{tab:ICmean_med}.

\subsection{Estimating the parameters of a Weibull distribution}\label{sec5.2}

The Weibull distribution with shape parameter $\beta>0$ and scale parameter $\eta>0$ has the density
$$f(x)=\frac{\beta}{ \eta^\beta} x^{\beta -1} \e^{-(x/\eta)^\beta},\quad x>0.$$
Based on a sample of $n$ independent realisations, many estimators of $\beta$ and $\eta$ are available (see \cite{johnson}). We consider the following three standard methods:
\begin{itemize}
\item the maximum likelihood estimator (ML)  is the solution of  the system 
$$\frac n \beta + \sum_{i=1}^n \log(x_i)- n \frac{\sum_{i=1}^n x_i^\beta \log(x_i)} {\sum_{i=1}^n x_i^\beta}=0,\quad \eta=\left(\frac 1 n \sum_{i=1}^n x_i^\beta\right)^{1/\beta}.$$

\item the method of moments (MM), based on the two first moments, reduces to solve:
$$\frac{s_n^2}{\overline x_n^2}=\frac{\Gamma(1+2/\beta)}{\Gamma(1+1/\beta)^2}-1,\quad \eta=\frac{\overline x_n}{\Gamma(1+1/\beta)},$$
where $\overline x_n$ and $s_n$ denote the empirical sample mean and the unbiased sample variance.

\item the ordinary least squares method (OLS) is based on the fact that for any $x>0$, $\log(-\log(1-F(x))=\beta \log(x) -\beta\log\eta$,  where $F$ denotes the cumulative distribution function of the Weibull distribution.   More precisely, denoting $x_{(1)},\dots,x_{(n)}$ the ordered sample, an estimation of $\beta$ and $\eta$ is  deduced from the simple  linear regression of $(\log(-\log(1-F(x_{(i)})))_{i=1\dots n}$ on $(\log x_{(i)})_{i=1\dots n}$, where according to the  "mean rank" method $F(x_{(i)})$ may be estimated by $i/(n+1)$. This fitting method is popular in the engineer community (see \cite{abernethy2006}): the estimation of $\beta$ simply corresponds to the slope in a "Weibull plot". 

\end{itemize}

\begin{figure}[!htbp]%
  \centering 
  
    \includegraphics[scale=.48]{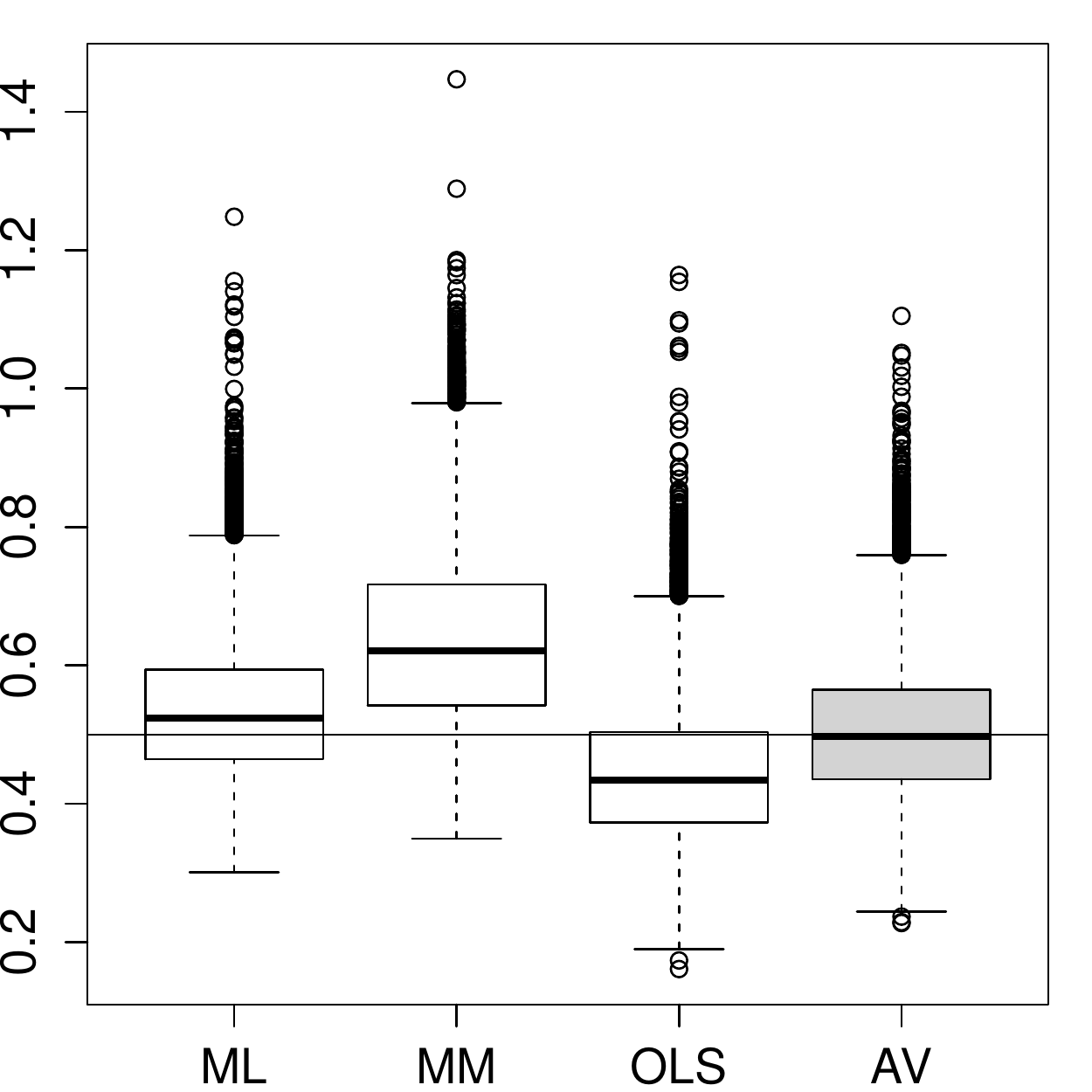}
  \quad
    \includegraphics[scale=.48]{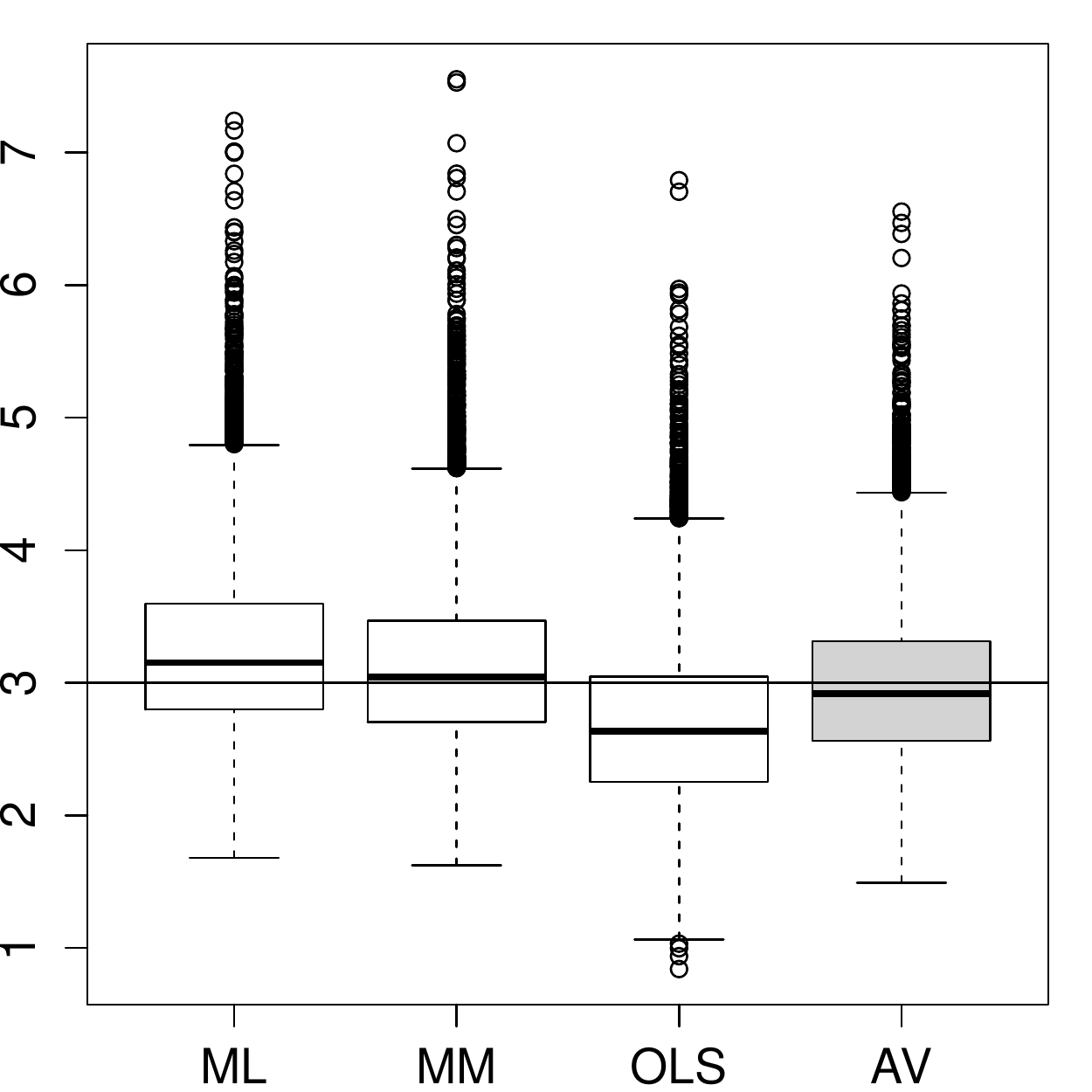}
  \caption{Repartition of $\hat\beta_{ML}$, $\hat\beta_{MM}$, $\hat\beta_{OLS}$ and $\hat\beta_{AV}$ (from left to right) based on $10^4$ replications of a sample of size $n=20$ from  a Weibull distribution with  $\beta=0.5$ (left), $\beta=3$ (right) and $\eta=10$.}
  \label{fig:weibull}
\end{figure}

The performances of these three estimators are variable, depending on the value of the parameters and the sample size. In particular, no one is uniformly better than the others, see Figure~\ref{fig:weibull} for an illustration.\\

Let us now consider the averaging of these estimators. In the setting of the previous sections, we have $d=2$ parameters to estimate and $k_1=3$, $k_2=3$ initial estimators of each are available. The averaging over the maximal constraint set $\Lambda_{\max}$ demands to estimate the $6\times 6$ MSE matrix $\Sigma$, that involves $21$ unknown values. The Weibull distribution is often used to model lifetimes, and typically only a low number of observations are available to estimate the parameters. As a consequence averaging over $\Lambda_{\max}$ of the $6$ initial estimators above could be too demanding. Moreover, between the two parameters $\beta$ and $\eta$, the shape parameter $\beta$ is often the most important to identify, as  it characterizes for instance the type of failure rate in reliability engineering. For these reasons, we choose to average the three estimators of $\beta$ presented above, $\hat\beta_{ML}$, $\hat\beta_{MM}$ and $\hat\beta_{OLS}$, and to consider only one estimator of $\eta$: $\hat\eta_{ML}$ (where $\hat\beta_{ML}$ is used for its computation). The averaging over $\Lambda_{\max}$ of these $4$ estimators has three consequences: First, the number of unknown values in the MSE matrix is reduced to $10$. Second, the averaging estimator of $\beta$ depends only on  $\hat\beta_{ML}$, $\hat\beta_{MM}$ and $\hat\beta_{OLS}$, because $\hat\eta_{ML}$ has a zero weight from \eqref{cond}. This means that we actually implement a component-wise averaging for $\beta$. Third, the averaging estimator of $\eta$ equals $\hat\eta_{ML}$ plus some linear combination of $\hat\beta_{ML}$, $\hat\beta_{MM}$ and $\hat\beta_{OLS}$ where the weights sum to zero. This particular situation will allow us to see if $\hat\eta_{ML}$ can be improved by exploiting the correlation with the estimators of $\beta$, or if it is deteriorated.\\

So we have $d=2$, $k_1=3$, $k_2=1$, $\mathbf T_1=(\hat\beta_{ML}, \hat\beta_{MM}, \hat\beta_{OLS})^\top $,  $\mathbf T_2=\hat\eta_{ML}$ and the averaging estimator over $\Lambda_{\max}$ is given by \eqref{ag_max}, denoted by $(\hat\beta_{AV},\hat\eta_{AV})^\top $. The matrix $\Sigma$ is estimated by Monte Carlo simulations: Starting from initial estimates  $\hat\beta_0$, $\hat\eta_0$, we simulate $B$ samples of size $n$ of a Weibull distribution with parameters $\hat\beta_0$, $\hat\eta_0$. Then the four estimators are computed, which gives $\hat\beta^{(b)}_{ML}$, $\hat\beta^{(b)}_{MM}$, $\hat\beta^{(b)}_{OLS}$ and $\hat\eta^{(b)}_{ML}$, for $b=1, \dots, B$,  and each entry of $\Sigma$ is estimated by its empirical counterpart. For instance the estimation of $\E(\hat\beta_{ML}-\beta)(\hat\beta_{MM}-\beta)$ is $(1/B)\sum_{b=1}^B (\hat\beta^{(b)}_{ML}-\hat\beta_0)(\hat\beta^{(b)}_{MM}-\hat\beta_0)$. 
In our simulations, we chose $\hat\beta_0$ as the mean of $\mathbf T_1$ and $\hat\eta_0=\hat\eta_{ML}$. Note that $\Sigma$ having a parametric form ensures that $(\hat\beta_{AV},\hat\eta_{AV})^\top $ is asymptotically as efficient as the oracle, as explained in Section~\ref{sec:asymptotic}.\\

Table~\ref{tab:beta} gives the MSE, estimated from $10^4$ replications,  of each estimator of $\beta$, for $n=10,20,50$, and for $\beta=0.5, 1, 2, 3$, $\eta=10$, where for each replication  $B=1000$.  The averaging estimator has by far the lowest MSE, even for small samples. As an illustration, the repartition of each estimator, for $n=20$ and $\beta=0.5,3$, is represented in Figure~\ref{fig:weibull}.

Table~\ref{tab:eta} shows the MSE for $\hat\eta_{ML}$ and $\hat\eta_{AV}$ where only estimators of $\beta$ were used in attempt to improve $\hat\eta_{ML}$ by averaging. The performances of both estimators are similar, showing that the information coming from $\mathbf T_1$ did not help significantly improving $\hat\eta_{ML}$. On the other hand, the estimation of these (almost zero) weights might have deteriorated $\hat\eta_{ML}$, especially for small sample sizes. This did not happen. 

Finally, the empirical rate of coverage of the asymptotic confidence intervals based on $\hat\beta_{AV}$ and $\hat\eta_{AV}$ is given in Table~\ref{ic:weibull}, showing that it is not far from the nominal level $95\%$, even for the small sample sizes considered in this simulation. On the other hand, the length of these intervals are by construction smaller than the length of similar confidence intervals based on the initial estimators.

\begin{table}[htbp]
\begin{center}
\resizebox{\textwidth}{!} {
\begin{tabular}{lcccccccccccc}
\hline
& \multicolumn{ 4}{c}{$n=10$} & \multicolumn{ 4}{c}{$n=20$} & \multicolumn{ 4}{c}{$n=50$} \\ 
 & ML & MM & OLS & AV \hspace{0.15cm} & \hspace{0.15cm} ML & MM & OLS & AV \hspace{0.15cm} & \hspace{0.15cm} ML & MM & OLS & AV \\ \hline
\multirow{2}{*}{$\beta=0.5$}  & 35.53 & 76.95 & 24.41 & 25.27 \hspace{0.15cm} & \hspace{0.15cm} 12.06 & 35.57 & 13.74 & 10.5 \hspace{0.15cm} & \hspace{0.15cm} 3.7 & 14.19 & 6.04 & 3.52 \\ 
 & (0.91) & (1.27) & (0.40) & (0.64) \hspace{0.15cm} & \hspace{0.15cm} (0.26) & (0.52) & (0.19) & (0.19) \hspace{0.15cm} & \hspace{0.15cm} (0.07) & (0.20) & (0.08) & (0.06) \\ \hline
\multirow{2}{*}{$\beta=1$}  & 152.4 & 131.6 & 98.1 & 85.5 \hspace{0.15cm} & \hspace{0.15cm} 49.2 & 53.6 & 54.2 & 36.9  \hspace{0.15cm} & \hspace{0.15cm} 14.4 & 19.3 & 23.9 & 12.8 \\ 
 & (3.8) & (3.1) & (1.5) & (1.7) \hspace{0.15cm} & \hspace{0.15cm} (1.1) & (1.1) & (0.7) & (0.7) \hspace{0.15cm} & \hspace{0.15cm} (0.2) & (0.2) & (0.2) & (0.2) \\ \hline
\multirow{2}{*}{$\beta=2$} & 596.4 & 444.6 & 399.4 & 355.5 \hspace{0.15cm} & \hspace{0.15cm} 194.5 & 164.5 & 218 & 163.3 \hspace{0.15cm} & \hspace{0.15cm} 57.9 & 53.9 & 94.8 & 54.3 \\ 
 & (14.4) & (11.9) & (6.3) & (6.7) \hspace{0.15cm} & \hspace{0.15cm} (3.8) & (3.3) & (2.8) & (2.7) \hspace{0.15cm} & \hspace{0.15cm} (1.0) & (0.9) & (1.3) & (0.9) \\ \hline
\multirow{2}{*}{$\beta=3$} & 1369 & 1080 & 905 & 770 \hspace{0.15cm} & \hspace{0.15cm} 452 & 394 & 486 & 343 \hspace{0.15cm} & \hspace{0.15cm} 128 & 122 & 211 & 120 \\ 
 & (34.6) & (29.7) & (14.6) & (18.1) \hspace{0.15cm} & \hspace{0.15cm} (9.8) & (8.9) & (6.7) & (6.2) \hspace{0.15cm} & \hspace{0.15cm} (2.2) & (2.0) & (2.7) & (1.9) \\ \hline
\end{tabular}}
\end{center}
\caption{Monte Carlo estimation of the MSE of $\hat\beta_{ML}$, $\hat\beta_{MM}$, $\hat\beta_{OLS}$ and $\hat\beta_{AV}$, based on  $10^4$ replications of a sample of size $n=10,\, 20,\, 50$ from a Weibull distribution with parameters $\beta=0.5,\, 1,\, 2,\, 3$ and $\eta=10$. The standard deviation of the MSE estimations are given in parenthesis. Each entry has been multiplied by 100 for ease of presentation.}
\label{tab:beta}
\end{table}

%
%
%
%

\begin{table}[htbp]
\begin{center}

\begin{tabular}{lcccccc}

 \hline
& \multicolumn{ 2}{c}{$ \hspace{0.8cm} n=10  \hspace{1.2cm}$} & \multicolumn{ 2}{c}{$ \hspace{0.8cm}n=20 \hspace{1.2cm}$} & \multicolumn{ 2}{c}{$ \hspace{0.8cm} n=50 \hspace{1.2cm}$} \\ 

 & ML \hspace{-0.4cm} & \hspace{-0.4cm} AV & ML \hspace{-0.4cm} & \hspace{-0.4cm} AV & ML \hspace{-0.4cm} & \hspace{-0.4cm} AV \\ \hline
\multirow{2}{*}{$\beta=0.5$} & \multicolumn{ 1}{c}{60.59} \hspace{-0.4cm} & \hspace{-0.4cm} 55.61 & 25.96 \hspace{-0.4cm} & \hspace{-0.4cm} 24.56 & 9.57 \hspace{-0.4cm} & \hspace{-0.4cm} 9.38 \\ 
& (1.60) \hspace{-0.4cm} & \hspace{-0.4cm} (1.48) & (0.53) \hspace{-0.4cm} & \hspace{-0.4cm} (0.5) & (0.17) \hspace{-0.4cm} & \hspace{-0.4cm} (0.17) \\ \hline
\multirow{2}{*}{$\beta=1$} & \multicolumn{ 1}{c}{11.15} \hspace{-0.4cm} & \hspace{-0.4cm} 10.88 & 5.53 \hspace{-0.4cm} & \hspace{-0.4cm} 5.43 & 2.23 \hspace{-0.4cm} & \hspace{-0.4cm} 2.22 \\ 
& (0.18) \hspace{-0.4cm} & \hspace{-0.4cm} (0.17) & (0.08) \hspace{-0.4cm} & \hspace{-0.4cm} (0.08) & (0.03) \hspace{-0.4cm} & \hspace{-0.4cm} (0.03) \\ \hline
\multirow{2}{*}{$\beta=2$} & \multicolumn{ 1}{c}{2.71} & 2.74 & 1.36 & 1.37 & 0.55 & 0.56 \\ 
& (0.04) & (0.04) & (0.02) & (0.02) & (0.01) & (0.01) \\ \hline
\multirow{2}{*}{$\beta=3$} & \multicolumn{ 1}{c}{1.21} \hspace{-0.4cm} & \hspace{-0.4cm} 1.23 & 0.61 \hspace{-0.4cm} & \hspace{-0.4cm} 0.61 & 0.247 \hspace{-0.4cm} & \hspace{-0.4cm} 0.248 \\ 
 & (0.02) \hspace{-0.4cm} & \hspace{-0.4cm} (0.02) & (0.01) \hspace{-0.4cm} & \hspace{-0.4cm} (0.01) & (0.003) \hspace{-0.4cm} & \hspace{-0.4cm} (0.004) \\ \hline
\end{tabular}
\end{center}

\caption{Monte Carlo estimation of the MSE of $\hat\eta_{ML}$ and $\hat\eta_{AV}$, based  on $10^4$ replications of a sample of size $n=10,\, 20,\, 50$ from a Weibull distribution with parameters $\beta=0.5,\, 1,\, 2,\, 3$ and $\eta=10$. The standard deviation of the MSE estimations are given in parenthesis. Each entry has been multiplied by 100 for ease of presentation.}
\label{tab:eta}
\end{table}

%
%
%
%

\begin{table}[htbp]
\begin{center}
\resizebox{\textwidth}{!} {
\begin{tabular}{lcccccc}

 \hline
$\hspace{2cm} $ & \multicolumn{ 2}{c}{$ \hspace{0.8cm} n=10 \hspace{2cm} $} & \multicolumn{ 2}{c}{$ \hspace{0.8cm} n=20 \hspace{2cm} $} & \multicolumn{ 2}{c}{$ \hspace{0.8cm} n=50 \hspace{1.5cm} $} \\ 

 & $\hat\beta_{AV}$\hspace{-0.4cm} & \hspace{-0.4cm} $\hat\eta_{AV}$ & $\hat\beta_{AV}$\hspace{-0.4cm} & \hspace{-0.4cm}$\hat\eta_{AV}$& $\hat\beta_{AV}$\hspace{-0.4cm} & \hspace{-0.4cm}$\hat\eta_{AV}$ \\ \hline
\multirow{1}{*}{$\beta=0.5$} & \multicolumn{ 1}{c}{89.84}\hspace{-0.4cm} & \hspace{-0.4cm}87.48 & 93.43\hspace{-0.4cm} & \hspace{-0.4cm}90.01 & 95.41\hspace{-0.1cm} & \hspace{-0.1cm}93.07 \\ 
\hline
\multirow{1}{*}{$\beta=1$} & \multicolumn{ 1}{c}{87.25}\hspace{-0.4cm} & \hspace{-0.4cm}89.24 & 90.98\hspace{-0.4cm} & \hspace{-0.4cm}91.61 & 93.81\hspace{-0.1cm} & \hspace{-0.1cm}93.62 \\ 
\hline
\multirow{1}{*}{$\beta=2$} & \multicolumn{ 1}{c}{89.96}\hspace{-0.4cm} & \hspace{-0.4cm}91.36 & 91.77\hspace{-0.4cm} & \hspace{-0.4cm}93.39 & 93.09\hspace{-0.1cm} & \hspace{-0.1cm}94.20 \\ 
 \hline
\multirow{1}{*}{$\beta=3$} & \multicolumn{ 1}{c}{92.19}\hspace{-0.4cm} & \hspace{-0.4cm}92.38 & 92.86\hspace{-0.4cm} & \hspace{-0.4cm}93.83 & 94.25\hspace{-0.1cm} & \hspace{-0.1cm} 94.77 \\ 
 \hline
\end{tabular}}
\end{center}
\caption{Empirical rate of coverage (in $\%$) of the asymptotic $95\%$ confidence intervals based on  $\hat\beta_{AV}$ and $\hat\eta_{AV}$ for the parameters of a Weibull distribution, deduced from the same simulations as in Tables~\ref{tab:beta} and \ref{tab:eta}. }
\label{ic:weibull}
\end{table}

\newpage

\subsection{Estimation in a Boolean model}\label{sec:bool}

\begin{figure}[!htbp]%
  \centering 
  
   \includegraphics[scale=.19]{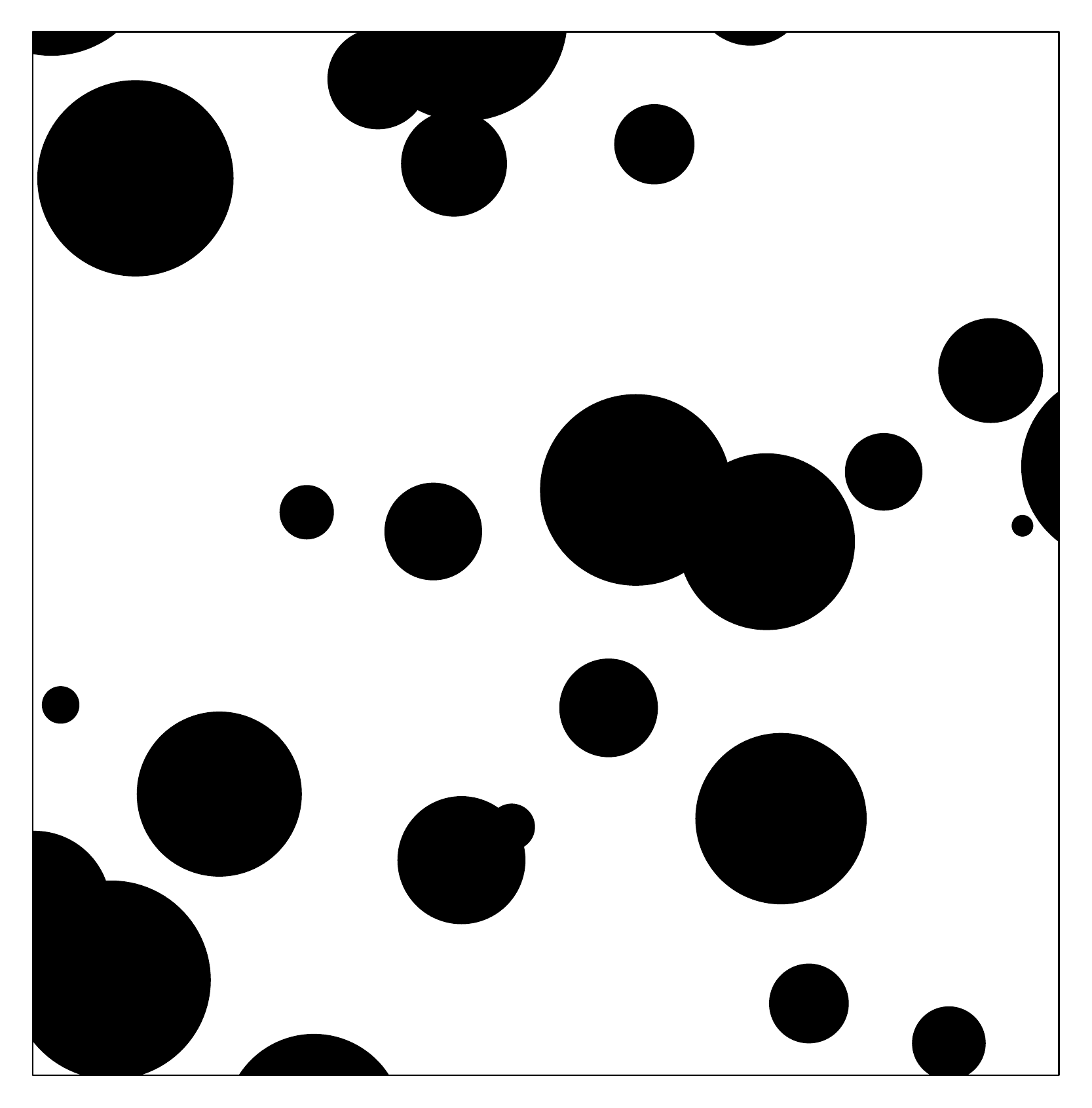}
    \includegraphics[scale=.19]{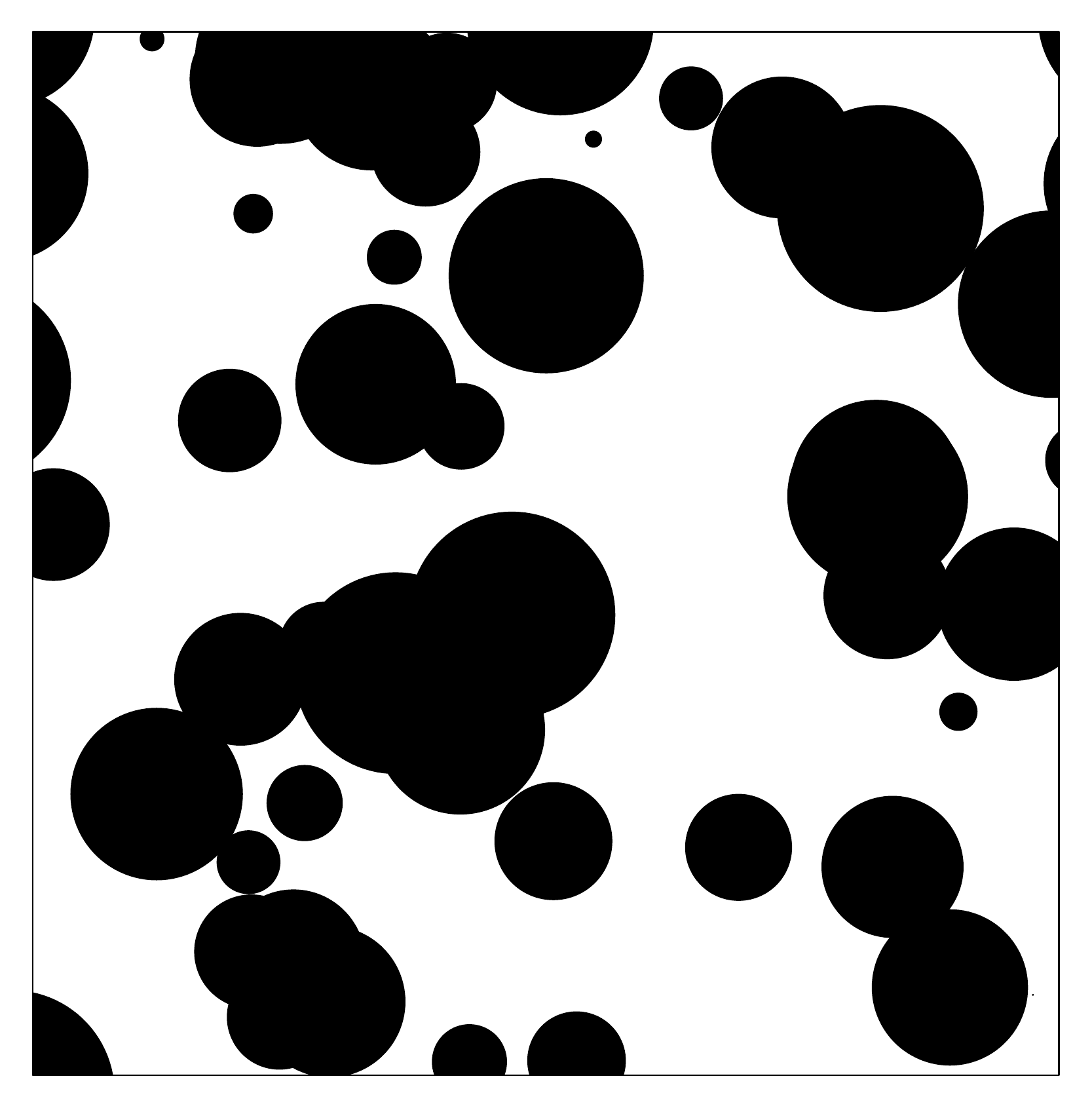}
       \includegraphics[scale=.19]{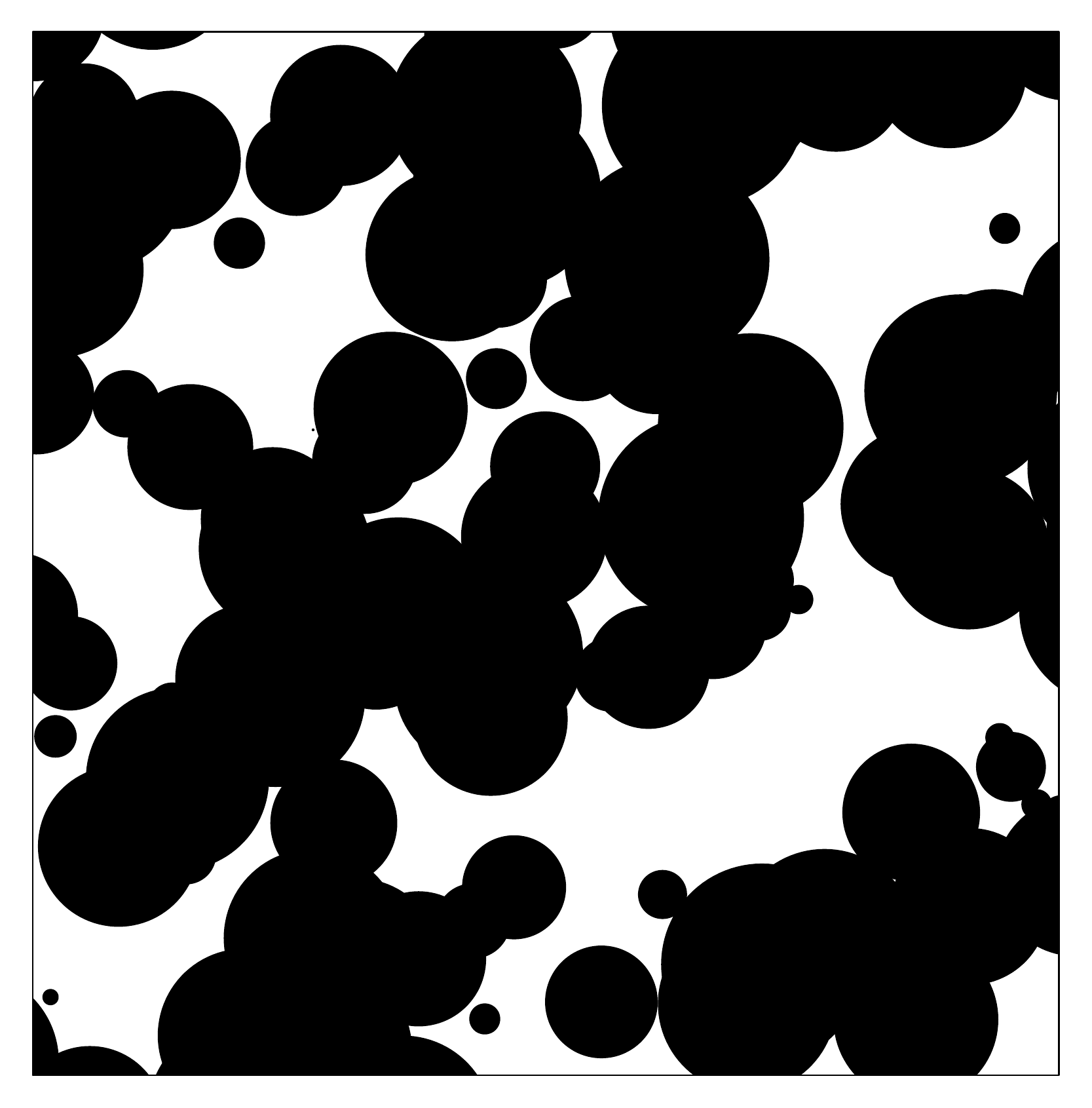}
    \includegraphics[scale=.19]{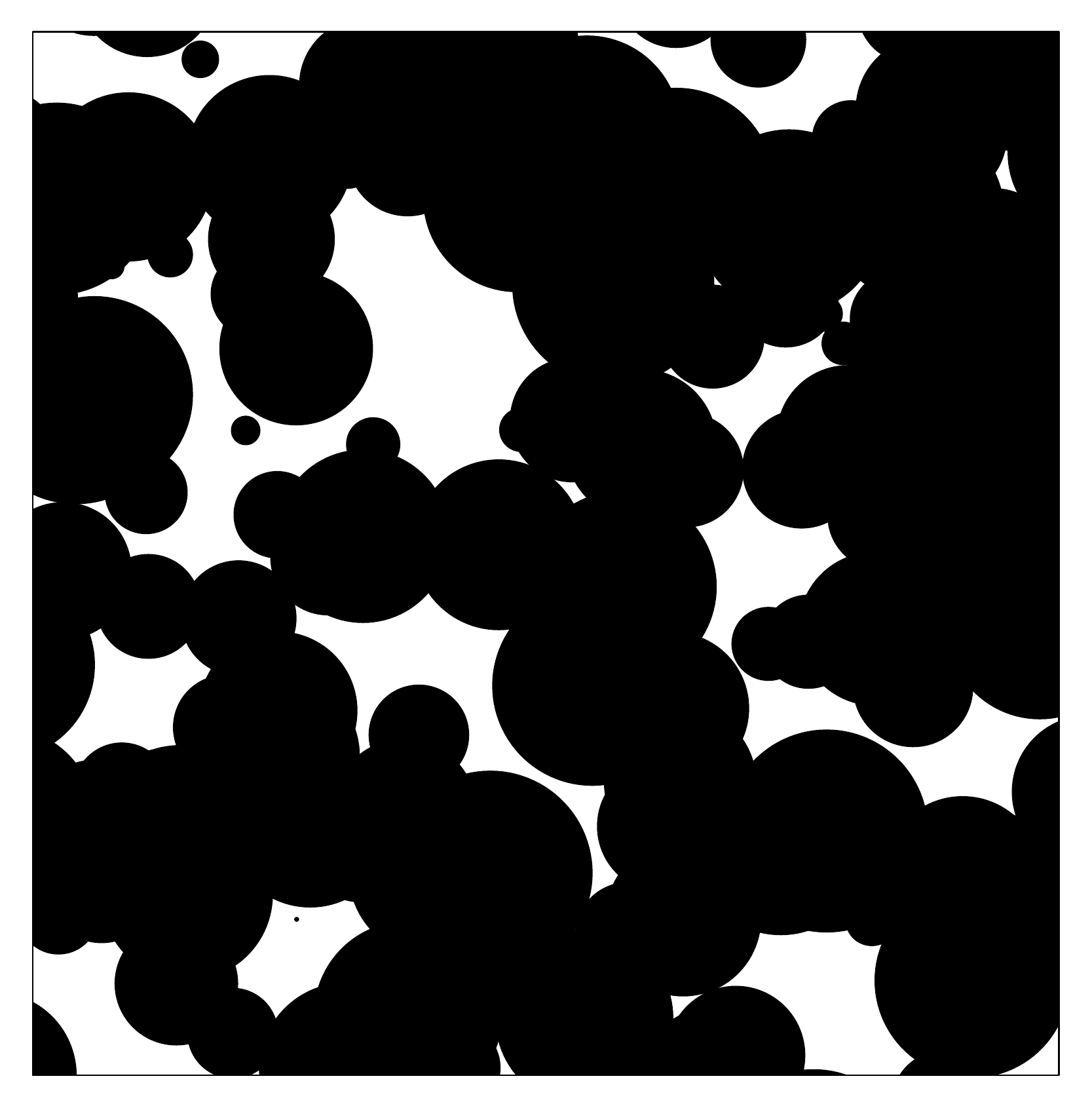}

  \caption{Samples from a Boolean model on $[0,1]^2$ with intensity, from left to right,  $\rho=25, 50, 100, 150$ and law of radii $B(1,\alpha)$ where $\alpha=1$.}
  \label{fig:bool}
\end{figure}

The Boolean model is the main model of random sets used in spatial statistics and stochastic geometry, see \cite{chiu2013}. It is a germ-grain model where, in the planar and stationary case, the germs come from a homogeneous Poisson point process on $\mathbb R^2$ with intensity $\rho$ and the grains are independent random discs, the radii of which are distributed according to a probability law $\mu$.  Figure~\ref{fig:bool} contains four realisations of a Boolean model on $[0,1]^2$ where $\rho=25, 50, 100, 150$ respectively and the law of the radii $\mu$ is the uniform  distribution over $[0,0.1]$. 
We assume in the following that $\mu$ is the beta distribution over $[0,0.1]$ with parameter $(1,\alpha)$, $\alpha>0$, denoted  by $B(1,\alpha)$, i.e. $\mu$ has  density $10\alpha\, (1-10 x)^{\alpha-1}$ on $[0,0.1]$. The simulations of  Figure~\ref{fig:bool} correspond to $\alpha=1$.\\

The estimation of  parameters $\rho$ and $\alpha$ from the observation of random sets as in Figure~\ref{fig:bool} is challenging, since the individual grains cannot be identified and likelihood-based inference is impossible. 
The standard method of inference, see \cite{Molchanov97}, is based on the following equations proved in \cite{weil1984}. They relate the expected area per unit area  $\mathcal A$ and the expected perimeter per unit area  $\mathcal P$ of the random set to the intensity $\rho$ and  the  two first moments of $\mu$, namely 
$$\mathcal A = 1- \exp(-\pi\rho  E_\mu(R^2)), \qquad \mathcal P = 2\pi \rho  E_\mu(R)  \exp(-\pi \rho E_\mu(R^2)),$$
where $R$ denotes a random variable with distribution $\mu$. Developing  $E_\mu(R)$ and $E_\mu(R^2)$ in terms of $\alpha$, we obtain the following estimates of $\alpha$ and $\rho$ :
$$\hat\alpha_1 = \frac{\mathcal P_{obs} }{10 (\mathcal A_{obs} - 1) \log(1-\mathcal A_{obs})} - 2, \qquad \hat\rho_1 = \frac {5\,(\hat\alpha_1+1)\mathcal P_{obs} }{\pi (1-\mathcal A_{obs})},$$
where $\mathcal A_{obs}$ and  $\mathcal P_{obs}$ denote the observed area and perimeter per unit area of the set.

An alternative procedure to estimate the intensity $\rho$ is based on the number of tangent points to the random set in a given direction. Let $u$ be a vector in $\mathbb R^2$. We denote by $N(u)$  the number of 
tangent points to the random set such that the associated tangent line is orthogonal to $u$ and the boundary of the set is convex in direction $u$. 
Considering $k$ distinct vectors $u_1,\dots,u_k$, an estimator of  $\rho$, studied in   \cite{molchanov1995}, is 
$$\hat\rho_2 = \frac{ \frac 1 k \sum_{i=1}^k N(u_i)}{|W| (1-\mathcal A_{obs})},$$
where $|W|$ denotes the area of the observation window. Although this estimator is consistent and asymptotically normal for $k=1$, it becomes more efficient as $k$ increases, see  \cite{molchanov1995}. In the following, we consider $k=100$ and the directions of $u_1,\dots,u_k$ are randomly drawn from an uniform distribution over $[0,2\pi]$.\\

Let us now consider the combination of the above estimators. In connection with the previous sections, we have $d=2$, $k_1=2$, $k_2=1$, $\mathbf T_1=(\hat\rho_1,\hat\rho_2)$ and $\mathbf T_2=\hat\alpha_1$. The averaging estimator over $\Lambda_{\max}$ is denoted by $(\hat\rho_{AV},\hat\alpha_{AV})$.  In this setting, we recall that $\hat\rho_{AV}$ is a linear combination of $\hat\rho_1$ and $\hat\rho_2$ where the weights sum to one, whereas $\hat\alpha_{AV}$ equals $\hat\alpha_1$ plus  a linear combination of $\hat\rho_1$ and $\hat\rho_2$ where the weights sum to zero. The weights are estimated according to \eqref{oracle_max}, where $\Sigma$ is obtained from Monte-Carlo simulations of the model with parameters $0.5 (\hat\rho_1+\hat\rho_2)$ and $\hat\alpha_1$ (see the previous section for more details). \\

Table~\ref{tab:bool} reports the MSE of each estimator, estimated from $10^4$ replications from a Boolean model with parameters $\rho=25, 50, 100, 150$ and $\alpha=1$. For each replication, 100 Monte-Carlo samples were used. The averaging estimators have better performances than the initial estimators. It is worth noticing the improvement of $\hat\alpha_1$ when it is corrected by $\hat\rho_1$ and $\hat\rho_2$ through $\hat\alpha_{AV}$. Though  this procedure might seem unnatural, the result is conclusive for this model. More simulations with other values of $\alpha$ (not reported in this paper) gave similar results.


\begin{table}[htbp]
\begin{center}
\begin{tabular}{lccccc}
\hline
  & $\hat \rho_1$ & $\hat \rho_2$ & $\hat \rho_{AV} $ & $ \hat\alpha_1$  &  $\hat\alpha_{AV}$ \\ \hline
\multirow{2}{*}{$\rho=25$}  &  34.15 & 14.63 & 14.60 & 8.09 & 6.70 \\ 
 & (0.55) & (0.22) & (0.22) & (0.15) & (0.13)  \\ \hline
\multirow{2}{*}{$\rho=50$}  &  131.63 & 47.41 & 45.65 & 4.69 & 3.24  \\ 
 & (2.26) & (0.72) & (0.67) & (0.067) & (0.048) \\ \hline
\multirow{2}{*}{$\rho=100$} &  949 & 272 & 223 & 5.70 & 2.29  \\ 
 & (21.8) & (4.9) & (3.6) & (0.086) & (0.034) \\ \hline
\multirow{2}{*}{$\rho=150$} &  7606 & 1656 & 1005 & 14.7 & 4.1 \\ 
 & (341) & (46.5) & (24.4) & (0.34) & (0.11)  \\ \hline
\end{tabular}
\end{center}
\caption{Monte Carlo estimation of the MSE of $\hat\rho_1$, $\hat\rho_2$, $\hat\rho_{AV}$ and $\hat\alpha_{1}$, $\hat\alpha_{AV}$ based on  $10^4$ replications of a Boolean model with parameters $\rho=25,\, 50,\, 100,\, 200$ and $\mu\sim B (1,\alpha)$ with $\alpha=1$. The standard deviation of the MSE estimations are given in parenthesis. The two last columns have been multiplied by 100 for ease of presentation.}
\label{tab:bool}
\end{table}

As explained  in Section~\ref{sec:asymptotic}, we can deduce from $\hat\rho_{AV}$ and $\hat\alpha_{AV}$ an asymptotic confidence interval without any further estimation. The length of this interval is smaller than the length of any similar confidence interval based on the initial estimators and  Table~\ref{tab:ICbool} reports the empirical rate of coverage of these intervals,   showing that it is close to the nominal level $95\%$.

%

\begin{table}[htbp]
\begin{center}
\begin{tabular}{lcccc}
\hline
& $ \hspace{0.2cm} \rho=25 \hspace{0.2cm} $ & $ \hspace{0.2cm} \rho=50 \hspace{0.3cm}  $ & $ \hspace{0.2cm} \rho=100 \hspace{0.2cm}  $ & $ \hspace{0.2cm} \rho=150 \hspace{0.2cm} $ \\ \hline
$\hat\rho_{AV}$ & 98.3 $\%$ &  97.6 $\%$ & 96.5 $\%$ &  93.4 $\%$ \\ \hline

$\hat\alpha_{AV}$ & 95.9 $\%$  &   94.3 $\%$  & 93.9 $\%$ &  94.9 $\%$\\ \hline
\end{tabular}
\end{center}
\caption{Empirical rate of coverage (in $\%$) of the asymptotic $95\%$  confidence intervals based on $\hat\rho_{AV}$ and $\hat\alpha_{AV}$  for the parameters of the Boolean model, deduced from the same simulations as in Table~\ref{tab:bool}. }
\label{tab:ICbool}
\end{table}

\subsection{Estimation of a quantile under misspecification}\label{quantiles}
In this section we consider a situation where we combine a non-parametric estimator with several parametric estimators coming from possibly misspecified models. In this setting, our main condition~\eqref{hypasymptotic} implying the asymptotic optimality of the average estimator is unlikely to hold and further investigations would be necessary to well understand the implications of a model misspecification. The following simulations nonetheless give an idea of the robustness of the averaging procedure. 

Specifically, given an iid sample $x_1,\dots,x_n$, we estimate the  $p$-th quantile of the unknown underlying  distribution by:
\begin{itemize}
\item the non-parametric estimator $\hat q_{NP} = x_{(\lfloor np\rfloor)}$; 
\item the parametric estimator associated to the Weibull distribution, i.e. $\hat q_W=F^{-1}_{\hat\alpha,\hat\beta} (p)$ where $F_{\alpha,\beta}$ denotes the cdf of the Weibull distribution with parameter $(\alpha,\beta)$ and $(\hat\alpha,\hat\beta)$ is  the MLE estimator;
\item the parametric estimator associated to the Gamma distribution;
\item the parametric estimator associated to the Burr distribution.
\end{itemize}
The three parametric models above have a different right-tail behavior: The Weibull distribution is not heavy-tailed when $\beta>1$, while the Gamma distribution is heavy-tailed but not fat-tailed and the Burr distribution is fat-tailed. \\

In this misspecified framework, we choose to use convex averaging, i.e.  the set of weights is given by \eqref{convex_weights}, and we denote by  $\hat q_{AV}$ the average estimator. The MSE matrix is estimated by bootstrap where for the initial estimator we take $\hat q_{NP}$.

Table~\ref{tab:quantiles} reports the mean squared error of each estimator when the ground truth distribution is either a Weibull distribution with parameter $(3,2)$, or the Gamma distribution with parameter $(3,2)$, or the Burr distribution with parameter $(2,1)$, or the standard lognormal distribution. Note that in the three first cases, one of the parametric estimators is well specified while the other parametric estimators are misspecified, and in the last situation all parametric estimators are misspecified. The table is concerned with  the estimation of the $p$-th quantile with $p=0.99$, based on $n=100$ and $n=1000$ observations. 
The mean squared errors are estimated from $10^4$ replications and the standard deviation of these estimations are given in parenthesis. Further simulations  have been conducted for other values of $p$, leading to the same conclusions as below, so we do not report them in this article. 

The average estimator outperforms the non-parametric estimator as soon as there is one well-specified parametric estimator in the initial collection, that is for the three first rows of Table~\ref{tab:quantiles}. When no parametric model is well-specified, as this is the case for the last row of Table~\ref{tab:quantiles}, then the average estimator has a similar mean squared error as $\hat q_{NP}$.

\begin{table}[htbp]
\begin{center}
\resizebox{\textwidth}{!} {
\begin{tabular}{lcccccccccc}
\hline
& \multicolumn{ 5}{c}{$\hspace{3cm} n=100\hspace{3cm} $} & \multicolumn{ 5}{c}{$\hspace{3cm} n=1000 \hspace{3cm} $}  \\ 
  & $\hat q_W$ & $\hat q_G$ & $\hat q_B $ & $\hat q_{NP}$  &  $\hat q_{AV} \hspace{0.28cm} $ & $\hspace{0.28cm}  \hat q_W$ & $\hat q_G$ & $\hat q_B $ & $\hat q_{NP}$  &  $\hat q_{AV}$ \hspace{-0.3cm} \\ \hline
\multirow{2}{*}{Weibull}  & 22 & 255 &  $1.10^5$ & 52 & 42 \hspace{0.28cm}  & \hspace{0.28cm}  2.1	& 234 & $1.10^5$ & 5.7	& 4.7 \hspace{-0.3cm} \\ 
 &(0.30) &	(2.0)	& ($405$)	& (0.7)	& (0.6) \hspace{0.28cm}  & \hspace{0.28cm} (0.03) &	(0.61)	& ($1.10^2$)	& (0.08)	& (0.07) \hspace{-0.3cm} \\ \hline
\multirow{2}{*}{Gamma}  &  3.6 & 1.8 &	$3.10^6$ & 	5.3 &	4.2\hspace{0.28cm} &\hspace{0.28cm} 1.7	& 0.18 &	$3.10^6$ & 	0.62 &	0.60  \hspace{-0.3cm} \\ 
 & (0.04) &	(0.02)	&  ($2.10^4$)	& (0.07)	& (0.06) \hspace{0.28cm}  & \hspace{0.28cm} (0.01) &	(0.003)	&  ($5.10^4$)	& (0.008)	& (0.008)
 \hspace{-0.3cm} \\ \hline
\multirow{2}{*}{Burr} &  15 &	18	 & 7 	& 24 &	16 \hspace{0.28cm}  & \hspace{0.28cm} 8.9 &	12.7	 &0.6	 &2.4 &	1.8
  \hspace{-0.3cm} \\ 
 & (0.14)&	(0.35)	 & (0.17)	& (1.46)	& (0.82) \hspace{0.28cm}  & \hspace{0.28cm} (0.04)&	(0.06)	 & (0.01)	& (0.04)	& (0.03) \hspace{-0.3cm}\\ \hline
\multirow{2}{*}{Lognormal} &  9.9 & 11.6 & 30.2 & 10.9 & 9.2 \hspace{0.28cm}  & \hspace{0.28cm}  7.29 & 9.87 & 13.39 & 1.38 & 1.38 \hspace{-0.3cm} \\ 
 & (0.08) & (0.07) & (0.51) & (0.24) & (0.13) \hspace{0.28cm}  & \hspace{0.28cm} (0.03) & (0.03) & (0.09) & (0.02) & (0.02) \hspace{-0.3cm} \\ \hline
\end{tabular}
}
\end{center}
\caption{Monte Carlo estimation of the MSE of $\hat q_W$, $\hat q_G$, $\hat q_B$, $\hat q_{NP}$ and $\hat q_{AV}$  when $p=0.99$, $n=100$ (left) and $n=1000$ (right),  based on  $10^4$ replications of a Weibull distribution (first row), a Gamma distribution (second row), a Burr distribution (third row) and a lognormal distribution (last row). The standard deviation of the MSE estimations are given in parenthesis. The first row has been multiplied by 1000  for ease of presentation.
}
\label{tab:quantiles}
\end{table}

\section{Appendix}

\subsubsection*{Proof of Theorem \ref{slutsky2multi}.} 
\noindent Since $\Lambda \subseteq \Lambda_{\max}$, we know that $\lambda^\top  \operatorname{J} = \text I$ for all $\lambda \in \Lambda$. Let $\mathbf S = \Sigma^{-\frac 1 2}  (\mathbf T - \operatorname{J} \theta ) $, we have
\begin{equation}\label{eqfrob} \Vert \hat \theta - \hat \theta^* \Vert^2 =  \Vert (\hat  \lambda - \lambda^* )^\top  (\mathbf T - \operatorname{J} \theta )\Vert^2 = \Vert (\hat  \lambda - \lambda^* )^\top  \Sigma^{\frac 1 2} \mathbf S \Vert^2 \leq \Vert (\hat  \lambda - \lambda^* )^\top  \Sigma^{\frac 1 2} \Vert_F^2 \ \Vert \mathbf S \Vert^2,
\end{equation}
where $\Vert A \Vert_F= \sqrt{\tr(A^\top A)}$ denotes the Frobenius norm of $A$. The map $\phi: \lambda \mapsto \tr(\lambda^\top  \Sigma \lambda)$ is coercive, and strictly convex by assumption. So, since $\Lambda$ is closed and convex, the minimum of $\phi$ on $\Lambda$ is reached at a unique point $\lambda^* \in \Lambda$. Moreover, we know that for $\lambda \in \Lambda$, $ \lambda^* + t(\lambda - \lambda^*)$ lies in $ \Lambda$ for all $ t \in [0,1]$, to which we deduce the optimality condition
$$ \lim_{t \to 0^+} \frac{\phi(\lambda^* + t(\lambda - \lambda^*)) - \phi(\lambda^*)} t = \tr \big[ \nabla \phi(\lambda^*)^\top (\lambda - \lambda^*) \big] = 2 \tr \big[ {\lambda^*}^\top  \Sigma ( \lambda - \lambda^*) \big] \geq 0, $$
for all $\lambda \in \Lambda$. It follows that
\begin{align}\label{eq2} \Vert (\hat  \lambda - \lambda^*) \Sigma^{\frac 1 2} \Vert_F^2 & = \tr ( \hat \lambda^\top  \Sigma \hat \lambda) - \tr ( {\lambda^*}^\top  \Sigma \lambda^*)   \nonumber- 2 \tr \big[ {\lambda^*}^\top  \Sigma ( \hat\lambda - \lambda^*) \big] \\
& \leq \tr ( \hat \lambda^\top  \Sigma \hat \lambda) - \tr ( {\lambda^*}^\top  \Sigma \lambda^*). 
\end{align}
By construction of $\hat \lambda$, we know that $\tr (\hat \lambda^\top  \hat \Sigma \hat \lambda) \leq  \tr({\lambda^*}^\top  \hat\Sigma {\lambda^*})$, yielding
\begin{eqnarray*} \tr ( \hat \lambda^\top  \Sigma \hat \lambda) - \tr ( {\lambda^*}^\top  \Sigma \lambda^*) & \leq  & \tr (  \hat \lambda^\top  \Sigma \hat \lambda) - \tr ( \hat \lambda^\top  \hat \Sigma \hat \lambda) + \tr ( {\lambda^*}^\top  \hat\Sigma {\lambda^*}) - \tr ( {\lambda^*}^\top  \Sigma \lambda^*) \\
& \leq & \tr ( {\hat \lambda}^\top  \hat \Sigma \hat \lambda ) \ \delta_\Lambda(\Sigma \vert \hat \Sigma) + \tr ({\lambda^*}^\top  \Sigma \lambda^*) \ \delta_\Lambda(\hat \Sigma \vert \Sigma) \\
& \leq &  \big[ \tr ( {\hat \lambda}^\top  \hat \Sigma \hat \lambda ) +  \tr ({\lambda^*}^\top  \Sigma \lambda^*) \big]  \delta_\Lambda( \hat \Sigma , \Sigma)
\end{eqnarray*}
where $\delta_\Lambda(A \vert B)$ and $\delta_\Lambda(A , B)$ are defined in Section \ref{errorbound}. Now using that $\tr ( {\hat \lambda}^\top  \hat \Sigma \hat \lambda ) \leq \tr ( {\lambda^*}^\top  \hat \Sigma \lambda^* )$ and
\begin{eqnarray*} \tr ( {\lambda^*}^\top  \hat \Sigma \lambda^* ) & = & \tr ({\lambda^*}^\top  \Sigma \lambda^*) + \big[\tr ( {\lambda^*}^\top  \hat \Sigma \lambda^*) - \tr ({\lambda^*}^\top  \Sigma \lambda^*)\big] \\
&  \leq & \tr ({\lambda^*}^\top  \Sigma \lambda^*) \big[ 1 + \delta_\Lambda (\hat \Sigma, \Sigma) \big],
\end{eqnarray*}
we obtain
\begin{equation}\label{eq3} \tr ( \hat \lambda^\top  \Sigma \hat \lambda) - \tr ( {\lambda^*}^\top  \Sigma \lambda^*) \leq \tr ({\lambda^*}^\top  \Sigma \lambda^*) \ \big[ 2 \delta_\Lambda (\hat \Sigma, \Sigma) + \delta_\Lambda (\hat \Sigma, \Sigma)^2 \big]. \end{equation}
Recall that $\tr ({\lambda^*}^\top  \Sigma \lambda^*) = \inf_{\lambda \in \Lambda} \ \mathbb E \Vert \lambda^\top  \mathbf T - \theta \Vert^2 = \mathbb E \Vert \hat \theta^* - \theta \Vert^2$, the result follows from \eqref{eqfrob}, \eqref{eq2} and \eqref{eq3}.\qed

\subsubsection*{Proof of Corollary \ref{cortheoric}.} 
 \noindent Write for $\epsilon >0$,
$$\Vert \hat \theta - \theta \Vert^2 \leq (1+\epsilon) \Vert \hat \theta^* - \theta \Vert^2 + (1+\epsilon^{-1}) \Vert \hat \theta - \hat \theta^* \Vert^2.$$ 
Applying Theorem \ref{slutsky2multi}, we get
\begin{equation}\label{eq8} \Vert \hat \theta - \theta \Vert^2 \leq (1+\epsilon) \Vert \hat \theta^* - \theta\Vert ^2 + (1+ \epsilon^{-1})\mathbb E\Vert \hat \theta^* - \theta \Vert^2 \ \left( \tilde \delta_\Lambda (\hat \Sigma^{}, \Sigma^{})  \ \Vert \mathbf S \Vert^2 \right),  \end{equation}
and the result follows by taking the expectation. \qed

\begin{lemma}\label{lem} Let $A$, $B$ be two positive definite matrices of order $k$. For any non-empty set $\Lambda$ that does not contain $0$,
$$ \delta_\Lambda(A, B) \leq \Vert \hspace{-0.04cm} \vert A B^{- 1}  - B A^{- 1}  \Vert \hspace{-0.04cm} \vert,    $$
where $\Vert \hspace{-0.04cm} \vert A \Vert \hspace{-0.04cm} \vert = \sup_{\Vert x \Vert_F = 1} \ \Vert A x \Vert_F$ stands for the operator norm. 
\end{lemma}
\noindent \textit{Proof.} By symmetry, it is sufficient to show that the result holds for $ \delta_\Lambda(A \vert B)$. We have
$$ \delta_\Lambda(A \vert B) = \sup_{\lambda \in \Lambda} \ \frac{\vert \tr[\lambda^\top  (B-A) \lambda] \vert}{\tr(\lambda^\top  B \lambda)} \leq \sup_{\lambda \neq 0}  \ \frac{\vert \tr[\lambda^\top  (B-A) \lambda] \vert}{\tr(\lambda^\top  B \lambda)}.  $$
By Cauchy-Schwarz inequality,
\begin{align} \vert \tr [\lambda^\top  (B - A) \lambda] \vert & = \big| \tr \big[ \lambda^\top  B^{\frac 1 2} \ (\text I - B^{- \frac 1 2} A B^{-\frac 1 2} ) \ B^{\frac 1 2} \lambda \big] \big| \nonumber \\ 
& \leq \Vert B^{\frac 1 2} \lambda \Vert_F \Vert (\text I - B^{- \frac 1 2} A B^{-\frac 1 2} ) \ B^{\frac 1 2} \lambda \Vert_F \nonumber \\
& \leq \Vert \hspace{-0.04cm} \vert \text I - B^{- \frac 1 2} A B^{-\frac 1 2} \Vert \hspace{-0.04cm} \vert \Vert B^{\frac 1 2} \lambda \Vert_F^2.
\end{align}
Recall that $\Vert B^{\frac 1 2} \lambda \Vert_F^2 = \tr(\lambda^\top  B \lambda)$, it follows
$$ \delta_\Lambda(A \vert B) \leq \Vert \hspace{-0.04cm} \vert \text I - B^{- \frac 1 2} A B^{-\frac 1 2} \Vert \hspace{-0.04cm} \vert. $$
Since the matrix $C=I - B^{- \frac 1 2} A B^{-\frac 1 2}$ is symmetric, it is diagonalizable in an orthogonal basis. In particular, denoting sp$(.)$  the spectrum, $\Vert \hspace{-0.04cm} \vert C  \Vert \hspace{-0.04cm} \vert=\sup_{t\in \operatorname{sp}(C)} |t|$. Finally, observe that $\text{sp}(C) = 1 - \text{sp}(B^{-\frac 1 2} A B^{- \frac 1 2}) = 1 - \text{sp}(AB^{-1})$, so that $AB^{-1}$  has positive eigenvalues and
$$ \Vert \hspace{-0.04cm} \vert \text I - B^{- \frac 1 2} A B^{-\frac 1 2} \Vert \hspace{-0.04cm} \vert = \sup_{t \in \operatorname{sp}(A B^{-1})} \ \vert 1 - t \vert \leq \sup_{t \in \operatorname{sp}(A B^{-1})} \ \vert t - \frac 1 t \vert  \leq \Vert \hspace{-0.04cm} \vert A B^{- 1}  - B A^{- 1}  \Vert \hspace{-0.04cm} \vert,   $$
ending the proof. \qed

\subsubsection*{Proof of Proposition \ref{cor}.}

\noindent By Lemma \ref{lem}, we know that $\delta_\Lambda (\hat \Sigma^{}_n, \Sigma^{}_n) = o_p(1)$ whenever $\hat \Sigma_n \Sigma_n^{- 1} \overset{p}{\longrightarrow} \text I$. Letting $\mathbf S_n = \Sigma_n^{- \frac 1 2} (\mathbf T_n - \operatorname{J} \theta)$, the fact that $\mathbb E \Vert \mathbf S_n \Vert^2 = k$ implies $\Vert \mathbf S_n \Vert^2 = O_p(1)$. Equation \eqref{eq8} holds for all $\epsilon >0$ so we can take $\epsilon = \epsilon_n$ such that $\epsilon_n \to 0$ and $\delta_\Lambda (\hat \Sigma^{}_n, \Sigma^{}_n)/\epsilon_n \overset{p}{\longrightarrow} 0$ as $n \to \infty$, yielding
$$ \Vert \hat \theta_n - \theta \Vert^2 \leq \Vert \hat \theta^*_n - \theta\Vert ^2 + \epsilon_n \Vert \hat \theta^*_n - \theta\Vert ^2 +  o_p(\alpha_n) = \Vert \hat \theta^*_n - \theta\Vert ^2 + o_p(\alpha_n).   $$
We shall now prove the second part of the proposition. Write,
\begin{equation} \hat \alpha_{n,j}^{-\frac 1 2} (\hat \theta_{n,j} - \theta_j) = \sqrt{\frac{\alpha_{n,j}}{\hat \alpha_{n,j}}} \ \alpha_{n,j}^{-\frac 1 2} \big[ (\hat \theta^*_{n,j} - \theta_j) + ( \hat \theta_{n,j} - \hat \theta^*_{n,j}) \big]. \nonumber \end{equation}
To prove the result, it suffices to show that $\alpha_{n,j}^{-\frac 1 2} \Vert \hat \theta_{n,j} - \hat \theta^*_{n,j} \Vert = o_p(1)$ and $\alpha_{n,j}/\hat \alpha_{n,j} \overset{p}{\longrightarrow} 1$. When $\Lambda$ is a cylinder, it is easy to see that the following holds
$$ \underline{\hat \lambda}_{n,j} = \arg \min_{\lambda \in  \Lambda_j} \ \lambda^\top \hat \Sigma_{n} \lambda \ \text{ and } \  \underline{\lambda}^*_{n,j} = \arg \min_{\lambda \in  \Lambda_j} \ \lambda^\top \Sigma_{n} \lambda, $$
where we recall $\Lambda_j = \{ \underline{\lambda}_j: \lambda \in \Lambda \}$. From the proof of Theorem~\ref{slutsky2multi}, we get
$$ \Vert \hat \theta_{n,j} - \hat \theta^*_{n,j} \Vert^2 \leq \alpha_{n,j} \ \big( 2 \delta_{\Lambda_j} (\hat \Sigma_n^{}, \Sigma_n^{}) + \delta_{\Lambda_j} (\hat \Sigma_n^{}, \Sigma_n^{} )^2 \big) \ \Vert \Sigma_n^{- \frac 1 2} (\mathbf T_n - \operatorname{J} \theta) \Vert^2.   $$
We deduce that $\alpha_{n,j}^{-\frac 1 2} (\hat \theta_{n,j} - \hat \theta^*_{n,j}) = o_p(1)$ in view of \eqref{hypasymptotic} and Lemma \ref{lem}. Now, remark that
$$  \frac{\alpha_{n,j}}{\hat \alpha_{n,j}}=  \frac{\underline{\lambda}_{n,j}^{*\top} \Sigma_n \underline{\lambda}^*_{n,j}}{\underline{\hat \lambda}_{n,j}^\top \hat \Sigma_n \underline{\hat \lambda}_{n,j}} \leq \frac{\underline{\hat \lambda}_{n,j}^\top \Sigma_n \underline{\hat \lambda}_{n,j}}{\underline{\hat \lambda}_{n,j}^\top \hat \Sigma_n \underline{\hat \lambda}_{n,j}} - 1 +1 \leq \delta_{\Lambda_j}(\hat \Sigma_n, \Sigma_n) +1.  $$
Similarly, 
$$ \frac{\hat \alpha_{n,j}}{\alpha_{n,j}} \leq \delta_{\Lambda_j}(\hat \Sigma_n, \Sigma_n) +1,$$
proving that  $\alpha_{n,j}/\hat \alpha_{n,j} \overset{p}{\longrightarrow} 1$ by the squeeze theorem.\qed

\section*{Acknowledgments}
 The authors are grateful to Ali Charkhi and Gerda Claeskens for fruitful discussion and to anonymous referees for numerous suggestions and comments which helped improve this paper.

\bibliographystyle{acm}
\bibliography{aggreg}

\begin{thebibliography}{10}

\bibitem{abernethy2006}
{\sc Abernethy, R.}
\newblock {\em The New {W}eibull Handbook: {R}eliability and Statistical
  Analysis for Predicting Life, Safety, Supportability, Risk, Cost and Warranty
  Claims}, fifth~ed.
\newblock Barringer \& Associates, 2006.

\bibitem{bates1969combination}
{\sc Bates, J.~M., and Granger, C.~W.}
\newblock The combination of forecasts.
\newblock {\em Operations Research 20}, 4 (1969), 451--468.

\bibitem{buckland1997}
{\sc Buckland, S.~T., Burnham, K.~P., and Augustin, N.~H.}
\newblock Model selection: an integral part of inference.
\newblock {\em Biometrics\/} (1997), 603--618.

\bibitem{MR2280619}
{\sc Bunea, F., Tsybakov, A.~B., and Wegkamp, M.~H.}
\newblock Aggregation and sparsity via {$l_1$} penalized least squares.
\newblock In {\em Learning theory}, vol.~4005 of {\em Lecture Notes in Comput.
  Sci.} Springer, Berlin, 2006, pp.~379--391.

\bibitem{MR2397610}
{\sc Bunea, F., Tsybakov, A.~B., and Wegkamp, M.~H.}
\newblock Sparse density estimation with {$\ell_1$} penalties.
\newblock In {\em Learning theory}, vol.~4539 of {\em Lecture Notes in Comput.
  Sci.} Springer, Berlin, 2007, pp.~530--543.

\bibitem{Catoni2004}
{\sc Catoni, O.}
\newblock Statistical learning theory and stochastic optimization, {E}cole
  d'{\'e}t{\'e} de {P}robabilit{\'e}s de {S}aint-{F}lour {XXXI}--2001.
\newblock {\em Lecture Notes in Mathematics 1851\/} (2004), 1--269.

\bibitem{Cesa-Lugosi}
{\sc Cesa-Bianchi, N., and Lugosi, G.}
\newblock {\em Prediction, learning, and games}.
\newblock Cambridge University Press, New York, 2006.

\bibitem{chiu2013}
{\sc Chiu, S.~N., Stoyan, D., Kendall, W.~S., and Mecke, J.}
\newblock {\em Stochastic geometry and its applications}, 3~ed.
\newblock John Wiley \& Sons, 2013.

\bibitem{MR3059085}
{\sc Dalalyan, A.~S., and Salmon, J.}
\newblock Sharp oracle inequalities for aggregation of affine estimators.
\newblock {\em Ann. Statist. 40}, 4 (2012), 2327--2355.

\bibitem{elliott2011averaging}
{\sc Elliott, G.}
\newblock Averaging and the optimal combination of forecasts.
\newblock Tech. rep., UCSD Working Paper, 2011.

\bibitem{MR0107925}
{\sc Graybill, F.~A., and Deal, R.~B.}
\newblock Combining unbiased estimators.
\newblock {\em Biometrics 15\/} (1959), 543--550.

\bibitem{halperin1961almost}
{\sc Halperin, M.}
\newblock Almost linearly-optimum combination of unbiased estimates.
\newblock {\em Journal of the American Statistical Association 56}, 293 (1961),
  36--43.

\bibitem{hansen2007least}
{\sc Hansen, B.~E.}
\newblock Least squares model averaging.
\newblock {\em Econometrica 75}, 4 (2007), 1175--1189.

\bibitem{hansen2012jackknife}
{\sc Hansen, B.~E., and Racine, J.~S.}
\newblock Jackknife model averaging.
\newblock {\em Journal of Econometrics 167}, 1 (2012), 38--46.

\bibitem{hjort2003frequentist}
{\sc Hjort, N.~L., and Claeskens, G.}
\newblock Frequentist model average estimators.
\newblock {\em Journal of the American Statistical Association 98}, 464 (2003),
  879--899.

\bibitem{johnson}
{\sc Johnson, N.~L., Kotz, S., and Balakrishnan, N.}
\newblock {\em Continuous univariate distributions. {V}ol. 1}, second~ed.
\newblock Wiley Series in Probability and Mathematical Statistics: Applied
  Probability and Statistics. John Wiley \& Sons Inc., New York, 1994.

\bibitem{MR1792783}
{\sc Juditsky, A., and Nemirovski, A.}
\newblock Functional aggregation for nonparametric regression.
\newblock {\em Ann. Statist. 28}, 3 (2000), 681--712.

\bibitem{MR2126899}
{\sc Keller, T., and Olkin, I.}
\newblock Combining correlated unbiased estimators of the mean of a normal
  distribution.
\newblock In {\em A festschrift for {H}erman {R}ubin}, vol.~45 of {\em IMS
  Lecture Notes Monogr. Ser.} Inst. Math. Statist., Beachwood, OH, 2004,
  pp.~218--227.

\bibitem{laplace}
{\sc Laplace, P.-S.~{\relax de}.}
\newblock {\em Th\'eorie analytique des probabilit\'es. {V}ol. {II}}.
\newblock \'Editions Jacques Gabay, Paris, 1995.
\newblock Reprint of the 1820 third edition (Book II) and of the 1816, 1818,
  1820 and 1825 originals (Supplements).

\bibitem{MR0264806}
{\sc Mehta, J., and Gurland, J.}
\newblock On combining unbiased estimators of the mean.
\newblock {\em Trabajos de estad{\'\i}stica y de investigaci{\'o}n operativa
  20\/} (1969), 173--185.

\bibitem{molchanov1995}
{\sc Molchanov, I.~S.}
\newblock Statistics of the boolean model: from the estimation of means to the
  estimation of distributions.
\newblock {\em Advances in applied probability\/} (1995), 63--86.

\bibitem{Molchanov97}
{\sc Molchanov, I.~S.}
\newblock {\em Statistics of the Boolean Model for Practitioners and
  Mathematicians}.
\newblock Wiley, Chichester, 1997.

\bibitem{moral2010model}
{\sc Moral-Benito, E.}
\newblock Model averaging in economics: An overview.
\newblock {\em Journal of Economic Surveys\/} (2013), 1--30.

\bibitem{nemirovski2000}
{\sc Nemirovski, A.}
\newblock Topics in {N}on-{P}arametric {S}tatistics, {E}cole d'{\'e}t{\'e} de
  {P}robabilit{\'e}s de {S}aint-{F}lour {XXVIII}--1998.
\newblock {\em Lecture Note in Mathematics 1738\/} (2000).

\bibitem{nocedal2006}
{\sc Nocedal, J., and Wright, S.}
\newblock Numerical optimization.
\newblock {\em Springer, New York\/} (2006).

\bibitem{raftery1997bayesian}
{\sc Raftery, A.~E., Madigan, D., and Hoeting, J.~A.}
\newblock Bayesian model averaging for linear regression models.
\newblock {\em Journal of the American Statistical Association 92}, 437 (1997),
  179--191.

\bibitem{rigollettsybakov}
{\sc Rigollet, P., and Tsybakov, A.}
\newblock Linear and convex aggregation of density estimators.
\newblock {\em Mathematical Methods of Statistics 16}, 3 (2007), 260--280.

\bibitem{silverman}
{\sc Silverman, B.~W.}
\newblock {\em Density estimation for statistics and data analysis}.
\newblock Monographs on Statistics and Applied Probability. Chapman \& Hall,
  London, 1986.

\bibitem{stiegler}
{\sc Stigler, S.~M.}
\newblock Laplace, {F}isher, and the discovery of the concept of sufficiency.
\newblock {\em Biometrika 60}, 3 (1973), 439--445.

\bibitem{timmermann2006forecast}
{\sc Timmermann, A.}
\newblock Forecast combinations.
\newblock In {\em Handbook of Economic Forecasting}, G.~Elliott, C.~Granger,
  and A.~Timmermann, Eds. North Holland, Amsterdam, 2006, pp.~135--196.

\bibitem{MR3225246}
{\sc Wang, Z., Paterlini, S., Gao, F., and Yang, Y.}
\newblock Adaptive minimax regression estimation over sparse {$\ell_q$}-hulls.
\newblock {\em J. Mach. Learn. Res. 15\/} (2014), 1675--1711.

\bibitem{wasserman2000bayesian}
{\sc Wasserman, L.}
\newblock Bayesian model selection and model averaging.
\newblock {\em Journal of mathematical psychology 44}, 1 (2000), 92--107.

\bibitem{weil1984}
{\sc Weil, W., and Wieacker, J.~A.}
\newblock Densities for stationary random sets and point processes.
\newblock {\em Advances in applied probability\/} (1984), 324--346.

\bibitem{MR1762904}
{\sc Yang, Y.}
\newblock Mixing strategies for density estimation.
\newblock {\em Ann. Statist. 28}, 1 (2000), 75--87.

\bibitem{yang2004aggregating}
{\sc Yang, Y.}
\newblock Aggregating regression procedures to improve performance.
\newblock {\em Bernoulli 10}, 1 (2004), 25--47.

\end{thebibliography}

\end{document}